\documentclass[12pt,a4paper]{article}

\usepackage{times} 
\usepackage[a4paper, lmargin=2cm, rmargin=2cm , tmargin=2.5cm , bmargin=2.25cm  ,footskip=1.25cm]{geometry} 
\usepackage{graphicx} 
\usepackage{subfig}
\usepackage[square,sort&compress,comma,numbers]{natbib} 
\usepackage{amsfonts}
\usepackage{amsmath}
\usepackage{amssymb}
\usepackage[hyphens]{url} 
\usepackage{soul}
\usepackage{fancyhdr} 
\usepackage[explicit]{titlesec} 
\usepackage{abstract} 
\usepackage{authblk} 
\usepackage{tabularx}
\usepackage{multirow} 
\usepackage{colortbl,color} 
\definecolor{Gray}{gray}{0.8}
\usepackage{booktabs}
\usepackage{array}
\usepackage[super]{nth} 
\usepackage{algorithm}
\usepackage[end]{algpseudocode}

\usepackage{booktabs}           

\newcommand{\fr}{\boldsymbol{f}}

\newcommand{\gr}{\boldsymbol{g}}

\newcommand{\nux}{\boldsymbol{\nu}} 

\newcommand{\Nr}{\boldsymbol{N}}

\newcommand{\xr}{\boldsymbol{x}}
\newcommand{\ur}{\boldsymbol{u}}
\newcommand{\er}{\boldsymbol{e}}
\newcommand{\yr}{\boldsymbol{y}}
\newcommand{\hrr}{\boldsymbol{h}}

\newcommand{\Kr}{\boldsymbol{K}}

\newcommand{\ys}{y}

\newcommand{\real}{\mathcal{R}} 
\newcommand{\reals}[1]{{\real}^{#1}}

\usepackage{afterpage}          

\usepackage[dvipsnames]{xcolor} 

\renewcommand{\footrulewidth}{4pt}

\renewcommand{\footrulewidth}{2pt}
\renewcommand{\footrule}{\hbox to\headwidth{\color{RoyalBlue}\leaders\hrule height \footrulewidth\hfill}}

\titleformat{\section}{\normalfont\bfseries}{\thesection}{1em}{\MakeUppercase{#1}}  

\titleformat{\subsection}{\normalfont\bfseries\small}{\thesubsection}{1em}{{#1}} 

\titleformat{\subsubsection}{\normalfont\small}{\thesubsubsection}{1em}{{#1}} 

\algrenewcommand\algorithmicend{\textbf{end}}
\algrenewtext{EndFor}{\algorithmicend}

\makeatletter
\newcommand{\thickhline}{ \noalign {\ifnum 0=`} \fi \hrule height 1pt \futurelet \reserved@a \@xhline }
\newcommand{\morethickhline}{ \noalign {\ifnum 0=`} \fi \hrule height 2pt \futurelet \reserved@a \@xhline }
\newcolumntype{"}{@{\hskip\tabcolsep\vrule width 1pt\hskip\tabcolsep}}
\makeatother

\newlength{\Oldarrayrulewidth}

\begin{document}

\title{\normalsize\normalfont\bfseries LAUNCHER ATTITUDE CONTROL BASED ON INCREMENTAL NONLINEAR\\DYNAMIC INVERSION: A FEASIBILITY STUDY TOWARDS\\FAST AND ROBUST DESIGN APPROACHES}
\date{}
\author[(1)]{Pedro Simpl\'icio\footnote{Corresponding author, email: \texttt{pedro.simplicio@ext.esa.int}}}
\author[(2)]{Paul Acquatella}
\author[(3)]{Samir Bennani}
\affil[(1)]{ \textit{Aurora Technology for the European Space Agency, Noordwijk, The Netherlands}}
\affil[(2)]{ \textit{DLR, German Aerospace Center, Oberpfaffenhofen, Germany}}
\affil[(3)]{ \textit{European Space Agency, Noordwijk, The Netherlands}}

\renewcommand\Authands{, }
\renewcommand{\Authfont}{\normalsize\normalfont \bfseries}
\renewcommand{\Affilfont}{\normalsize\normalfont}
\renewcommand{\abstractnamefont}{\bfseries\normalsize\MakeUppercase} 

\maketitle

\begin{abstract}
{\footnotesize{
The so-called "New Space era" has seen a disruptive change in the business models and manufacturing technologies of launch vehicle companies. However, limited consideration has been given to the benefits that innovation in control theory can bring; not only in terms of increasing the limits of performance but also reducing mission preparation or ``missionisation'' efforts. Moreover, there is a gap between the current state-of-practice that still relies on linear controls and other modern control techniques that could bring relevant improvements in launcher attitude control; this is the case for nonlinear control algorithms, especially those based on Nonlinear Dynamic Inversion (NDI). NDI is a technique that basically `cancels' the nonlinearities of a class of nonlinear systems, allowing for a single linear control law to be applied without the need for gain-scheduling across different operational points. Incremental NDI (INDI) is a variation of NDI that generates incremental commands and employs acceleration feedback to reduce model dependency, making it easier to design, and results in being more robust in closed-loop. While INDI has been applied successfully to several aerospace applications, its applicability to launch vehicles has not yet been adequately investigated. The objective of this paper is therefore to introduce and raise awareness of the INDI method among the launcher guidance, navigation, and control (GNC) community, showcasing its implementation on a representative launch ascent application scenario which highlights INDI's strengths and challenges. We present a new, practical approach for stability analysis of INDI for attitude control, and compare INDI with scheduled PD controllers with- and without angular acceleration estimates. Results show that, while INDI controllers are generally more sensitive to sensor noise and actuator delay than linear controllers, their potential benefits outweigh these limitations in terms of robustness and performance.
}}
\end{abstract}

\thispagestyle{fancy}

\section{Introduction} \label{sec:intro}
\subsection{Background and Motivation}
The space industry has undergone significant changes in recent years with the advent of the ``New Space era'' marked by disruptive changes in the business models, manufacturing technologies, and agile practices of launch vehicle companies; all aimed at minimising their production and operating costs in an ever more competitive market. However, limited attention has been given to the benefits of control theory innovation in this context despite the potential for such innovations to increase performance limits and reduce mission preparation (or ``missionisation'') efforts. Moreover, government-led developments of recent launchers such as Ares I and VEGA still use the same design approach of the Saturn V, i.e. linear controllers~\cite{SICE2020}. This approach relies on single channel-at-a-time tuning and ad--hoc gain-scheduling followed by extensive validation and verification (V\&V); these are in fact quite time- and cost-consuming processes.

In contrast to the approach presented above, the past few years have seen a growing interest in the application of artificial intelligence and machine learning methods for launcher GNC, but the industrial use of such data-driven/model-free methods remains limited by well-known issues related to training and certification of the algorithms on the full flight envelope of intended operation. In that sense, there is a clear gap between these strategies and the current state-of-practice, in which other techniques could bring relevant improvements; this is the case for nonlinear control algorithms, especially those based on Nonlinear Dynamic Inversion (NDI). On one hand, agile practices of New Space companies provide the ideal opportunity to explore the benefits of this type of design approach. On the other hand, a successful adoption of nonlinear launcher control will likely facilitate the augmentation with and transition to data-driven methods in the future. This is therefore our motivation and aim for this paper, to start bridging the gap between these two approaches while presenting a potential alternative based on incremental nonlinear control.

\subsection{Related Work}

In this paper we introduce briefly and focus on (Incremental) Nonlinear Dynamic Inversion (NDI) which is a control design method based on \textit{feedback linearisation}~\cite{slotine1990,khalil2002}; it basically consists on a nonlinear (state feedback) transformation that linearises the nominal system dynamics, and a linear part that imposes the desired closed-loop dynamics. Actually, NDI is a very well known and applied (nonlinear) control technique in the aerospace field, especially in aeronautics for various flight control applications~\cite{balas1996,smith1998,looye2001,lombaerts2008}. Successful implementation of NDI requires a match between the onboard model and the system model, and accurate knowledge of all nonlinearities, which is often not the case in reality; this results in poor robustness properties because they rely on exact availability of the system dynamics. This highlights the need for robustness in these methodologies, as the inner-loop of the control system is critical and can be compromised by model and sensor uncertainties, potentially affecting stability and performance. In this regard, alternative methods involving robustness and improvements of the method for NDI-based flight control applications were considered, among many others, 
in~\cite{bennani1998, smith2000,bacon2000a,bacon2000b,bacon2001}. \\

A successful technique that became popular in the recent years for aerospace applications is Incremental Nonlinear Dynamics Inversion (INDI). The concept using incremental nonlinear control was first developed in the late nineties and was initially focused on the `\textit{implicit}' dynamic inversion for DI-based flight control. The works of Smith, Bacon, and others laid the foundation for these developments~\cite{smith1998, bacon2000a}, for which the term `\textit{incremental}' is now more commonly used to describe this methodology as it better reflects the nature of these control laws~\cite{chen2008, sieberling2010, simplicio2013}. Those early studies further developed the incremental approach and, since then, it has been further elaborated theoretically and successfully applied in various high-performance systems including fault-tolerant control of aircraft subjected to sensor and actuator faults~\cite{lu2015a,lu2015b}, 
in practice for quadrotors using adaptive control~\cite{smeur2016b,smeur2016a}, in real flight tests of small (unmanned) and business jet (Cessna Citation II, PH-LAB) aircraft~\cite{vlaar2014msc,gron2018,twan2019}, but also for spacecraft attitude control~\cite{acquatella2012,acquatella2020,acquatella2022}. However, its applicability to launch and re-entry vehicles has not been fully investigated but only considered in~\cite{mooij2020,mooij2021,mooij2023}, and planned to be flight-tested in the upcoming `\textit{Reusability Flight Experiment (ReFEx)}' by DLR~\cite{rickmers2021}. 
These related works have demonstrated INDI's performance and robustness against aerodynamic model uncertainties and disturbance rejection for several aerospace vehicles; hence, the potential benefits of INDI are quite relevant for reusable launchers which have much tighter dynamical couplings between online-generated trajectory and attitude control during descent flight. Moreover, due to the nonlinear nature of INDI, it has been proven difficult to attain an analytical proof of stability which has been derived in~\cite{wang2019indi}. With this paper we aim for further close this gap towards the application of INDI for launchers with special focus on the ascent of a TVC-controlled launcher and also aim to present a new, practical approach for stability analysis of such INDI control laws applied for attitude control.

\subsection{Objectives and Outline}
It is therefore the objective of this study to introduce and raise awareness of the INDI technique among the launcher GNC community, to showcase its implementation on a representative application scenario, and to highlight its strengths and challenges in the face of the industrial state-of-practice. To achieve this, the paper provides a concise description of the NDI and INDI approach, followed by the detailed design and comparison of different control laws: linear, linear with angular acceleration feedback and INDI-based. 
Furthermore, the paper is also aimed to address the (mainly) two well-known challenges associated with the practical implementation of INDI-based control:
\begin{itemize}
\item \textbf{Sensitivity to sensor noise and actuator delay}.
By relying on angular acceleration and control input measurements/estimates, INDI controllers are generally more sensitive to sensor noise and actuator delay than classical controllers. 
To assess the severity of this challenge, the paper shows a comprehensive nonlinear simulation campaign with wind disturbances, uncertainties, as well as different levels of sensor noise and actuator delay. These simulations serve as a basis to analyse the sensitivity to sensor noise and actuator delay in comparison to more classical approaches and we showcase how to remediate or tackle these issues properly.
\item \textbf{Nonlinear stability analysis}.
The second challenge of INDI is that, due to its nonlinear nature, attaining an analytical proof of stability is not trivial~\cite{wang2019indi}. For this second challenge, the paper proposes a simple yet insightful linearisation-based approach to evaluate stability degradation related to an inexact feedback linearisation and to deviations from the control tuning conditions. This method provides a new way to analyse and evaluate stability analysis of the nonlinear controller using linear control techniques; since INDI is designed from the theory of feedback linearisation, this approach is very intuitive in the sense it provides a measure of degradation with respect to the feedback linearised plant and linear stability analysis can be performed. 
\end{itemize}

To demonstrate the benefits and challenges of the INDI approach, we showcase the method within an application scenario consisting of a launcher model during ascent flight while featuring attitude and lateral drift degrees-of-freedom, actuator dynamics, and moving-mass effects. All the controllers and filters are implemented at a sampling frequency that is compatible with current onboard capabilities (25 Hz).

The outline of this paper is as follows. 
A brief introduction to Nonlinear Dynamic Inversion (NDI) and Incremental NDI is presented in Sec.~2. 
Section~3 presents the modelling aspects of the launcher application in consideration and describes the simulator used for
the attitude control design and testing. Launcher attitude control designs including angular acceleration feedback
are presented in Sec.~4. Time-domain robust performance results and analysis of the obtained simulations comparing the controllers studied are presented in Sec.~5, while Sec.~6 presents the frequency-domain stability results and analysis. Conclusions are finally presented in Sec.~7.

\section{Basic principles of (Incremental) Nonlinear Dynamic Inversion} \label{sec:indi}

\subsection{Nonlinear Dynamic Inversion (NDI)} \label{sec:intro_ndi}
 
Without loss of generality, we consider a 
multiple-input and multiple-output
(MIMO) system whose number of inputs are equal to the
number of outputs in order to avoid control allocation and internal dynamics problems. Let's also assume momentarily that the nonlinear system can be described affine in the inputs as:
\begin{subequations}
	\begin{equation}
		\dot{\xr} = \fr(\xr) + \gr(\xr)\ur
	\end{equation}
	\begin{equation}
		\yr = \hrr(\xr)
	\end{equation}
\end{subequations}
where $\xr\in\reals{n}$ is the state vector,
$\ur\in\reals{m}$ is the control input vector, and
$\yr\in\reals{m}$ is the system output vector,
the functions $\fr(\xr)$ and $\hrr(\xr)$
are assumed to be smooth vector fields on $\reals{n}$, and
$\gr(\xr)\in\reals{n\times m}$ 
is a matrix whose columns are also assumed as smooth vector fields $\gr_j$.
For these systems, the vector of relative degree represents the number of differentiations of each output $y_i$, $i = 1,\ldots,m$,
that are needed for the input to appear~\cite{slotine1990,khalil2002}.
In this brief introduction to NDI we consider $\yr = \xr$ so that the relative degree of each of the outputs $y_i$ is one; for a detailed explanation of NDI for higher relative degrees including the transformation of the nonlinear system into a normal form decomposed into an external (input–output) part and an internal (unobservable) part, the reader is referred to~\cite{wang2019indi,acquatella2022}.

Nonlinear Dynamic Inversion (NDI) is a technique that aims to eliminate the nonlinearities present in a given nonlinear system, resulting in closed-loop dynamics that can be expressed in a linear form. To achieve this, the nonlinear system is inverted into a linear structure using state feedback, making it possible to apply conventional linear controllers. However, NDI has a significant disadvantage in that it relies on the fundamental assumption that the system model is known exactly, making it vulnerable to uncertainties. Additionally, NDI assumes that the system state is fully and accurately known, which can be challenging to achieve in practice. NDI involves applying the following input transformation~\cite{slotine1990,khalil2002}:
\begin{equation}\label{eq:input-transformation-MIMO}
	\ur_{{\text{cmd}}} =
	\gr^{-1}(\xr)\left(\nux - \fr(\xr)\right)
\end{equation}
which cancels all nonlinearities in closed-loop, 
and a simple linear input-output relationship 
between the new virtual control input 
$\nux$ and the output $\yr$ is obtained:
\begin{align}\label{eq:linear-MIMO}
	\dot{\yr} = \nux 
\end{align}
In addition to being linear, an interesting feature of this relationship is that it is also decoupled, meaning that the input $\nu_i$ only affects the output $\ys_i$. This property gives rise to the so-called ``decoupling control law'' to describe the input transformation in \eqref{eq:input-transformation-MIMO}, and the resulting linear system in \eqref{eq:linear-MIMO} is referred to as a ``single-integrator'' form. By utilising appropriate (linear, robust) control techniques, the single-integrator form in \eqref{eq:linear-MIMO} can be rendered exponentially stable. For instance,
the single-integrator can be made exponentially stable through the use of: 
\begin{align}\label{eq:virtual-control-law-error}
	\nux = \dot{\yr}_{\text{des}} = \dot{\yr}_{\text{cmd}} + \Kr_P\,\er
\end{align}
where $\nux = \dot{\yr}_{\text{des}}$ defines the desired dynamics for the output vector or control variables. The feedforward term for tracking is given by $\dot{\yr}_{\text{cmd}}$, while $ \er = {\yr}_{\text{cmd}} - {\yr}$ represents the error vector. Here, ${\yr}_{\text{cmd}}$ denotes the (smooth) desired output vector, which is (in this case, since relative degree is one) at least once differentiable. The gain matrix $\Kr_P\in\reals{m\times m}$ is used to ensure that the polynomials given by $s + K_{P_i}$ for $i = {1,\ldots,m}$, become Hurwitz. The diagonal elements $K_{P_i}$ of $\Kr_P$ are then selected accordingly. As a result of using \eqref{eq:virtual-control-law-error}, the desired error dynamics $\dot{e}_i + K_{P_i}\,e_i = 0$, become exponentially stable and decoupled, leading to $e_i(t)\rightarrow 0$ for $i={1,\ldots,m}$.

\subsection{Incremental Nonlinear Dynamic Inversion (INDI)} \label{sec:intro_indi}

Incremental nonlinear dynamic inversion (INDI) consists on the application of NDI to a system expressed in an incremental form~\cite{sieberling2010, simplicio2013,wang2019indi}. To obtain a system in incremental form, first we introduce a \textit{sufficiently small} time--delay $\lambda$ and define the following deviation variables
$\dot{\xr}_0:= \dot{\xr}(t-\lambda)$,
$\xr_0:= \xr(t-\lambda)$, and
$\ur_0:= \ur(t-\lambda)$,
which are the $\lambda$--time--delayed signals of the current 
state derivative $\dot{\xr}(t)$, state $\xr(t)$, and control $\ur(t)$,
respectively~\cite{acquatella2020}. Moreover, we will denote
$\Delta\dot{\xr} := \dot{\xr}-\dot{\xr}_0$,
$\Delta{\xr} := {\xr}-{\xr}_0$, and
$\Delta\ur := \ur-\ur_0$
as the incremental state derivative, 
the incremental state, and
the so--called incremental control input, 
respectively. Subsequently, we
consider a first-order Taylor series expansion of $\dot{\xr}$,
not in the geometric sense, 
but with respect to the newly introduced time--delay $\lambda$ as~\cite{sieberling2010, simplicio2013,wang2019indi,acquatella2020}:
\begin{equation}\label{eq:taylor-incremental}
    \begin{split}
        \dot{\xr}  
	= ~ & \dot{\xr}_0 + 
	\frac{\partial}{\partial\xr}
	\big[{\fr}(\xr) + {\gr}(\xr)\ur\big] \bigg|_{\substack{\xr=\xr_0 \\ \ur=\ur_0}}
	\Delta\xr 
	+ {\gr}(\xr_0)\Delta\ur 
	+ \textit{H.O.T} \\
	= ~ 
	& \dot{\xr}_0  
	+ {\gr}(\xr_0)\Delta\ur	
	+ \Nr(\xr, \lambda)	
    \end{split}
\end{equation}
with:
\begin{subequations}
	\begin{align}
		\dot{\xr}_0 
		= \fr(\xr_0) + \gr(\xr_0)\ur_0
	\end{align}
	\begin{align}
		\Nr(\xr, \lambda) & = \frac{\partial}{\partial\xr}
		\Big[\fr(\xr) + \gr(\xr)\ur\Big] 
		\Big|_{\substack{\xr=\xr_0 \\ \ur=\ur_0}}\Delta\xr 
		+ \textit{H.O.T}
	\end{align}
\end{subequations}
which represents a residual containing the Jacobian linearisation of the on-board model and the higher order terms ($\textit{H.O.T}$) of the series expansion.
Notice that the model--based control effectiveness  $\boldsymbol{g}(\xr_0)$ is sampled at the previous incremental time.
This means an approximate linearisation about the $\lambda-$delayed signals is performed \textit{incrementally}, and not with respect to a particular equilibrium or operational point of interest. Further, we consider the following \textsl{time-scale separation assumption}:\\

\textit{For a sufficiently small time-delay $\lambda$ and for any incremental 
	control input, it is assumed 
	that $\Delta{\xr}$ does not vary significantly during $\lambda$. 
	In other words, the input rate of change
	is much faster than the state rate of change:}
\begin{equation}\label{eq:eps-indi-tss}
	{\epsilon_{INDI_{\text{TSS}}}}(t) 
	\equiv \Delta{\xr} := {\xr}-{\xr}_0 \cong 0, ~\forall~\Delta\ur
\end{equation}
\textit{which leads to:}
\begin{align*}\label{eq:taylor-incremental}
	\dot{\xr}
	\cong ~ & 
	\dot{\xr}_0 
	+ \gr(\xr_0)\cdot\left(\ur-\ur_0\right)
	+ \underbrace{\Nr(\xr, \lambda)}_{\cong 0}
\end{align*}
\textit{or simply:}
\begin{equation}\label{eq:inc-odot-rep}
	\Delta\dot{\xr} \cong \gr(\xr_0)\cdot\Delta\ur
\end{equation}\\

This assumption, corroborated by the fact that 
the perturbation term $\Nr(\xr, \lambda)$ satisfies~\cite{wang2019indi}:
\begin{equation}\label{eq:res-wang-ok}
	\lim_{\lambda\rightarrow 0}
	\left\|
	\Nr(\xr, \lambda)
	\right\|_2
	\rightarrow 0,~\forall\,\xr
\end{equation}
implies that the nonlinear system dynamics in its incremental form is approximated at each time-step by the model-based control effectiveness $\gr(\xr_0)$. Finally, applying
NDI to the system based on the approximation~\eqref{eq:inc-odot-rep}  results in a relation between the incremental control input and the output of the system:
\begin{equation}\label{eq:incremental-second-relation}
	\ur = \ur_0 + \gr(\xr_0)^{-1}(\nux-\dot{\xr}_0)	.
\end{equation}
and noticing that while implementing this control law it will be required the availability of $\dot{\xr}_0$ and that the incremental input $\ur_0$ 
is obtained from the output of the actuators or estimated from an actuator dynamical model; recall it has been assumed that a commanded control is achieved \textit{sufficiently fast} in regards to the actuator dynamics. The total control command along with the obtained linearising control $\ur_0=\ur(t-\lambda)$ can be rewritten as:
\begin{equation}\label{eq:incremental-final}
	\ur(t) = \ur(t-\lambda) + \gr(\xr_0)^{-1}[\,\nux-\dot{\xr}(t-\lambda)].
\end{equation}
This improves the robustness of the closed-loop 
system as compared with conventional NDI since dependency on the accurate 
knowledge of the plant dynamics is reduced; more specifically, the dependency on accurate knowledge of the dynamic model in 
$\fr(\xr)$ is largely decreased. Therefore, the INDI control law design is more dependent on accurate measurements or accurate estimates of $\dot{\xr}_0$, the state derivatives, and $\ur_0$, the incremental control input, respectively. 
%

\section{Launcher model and simulator description} \label{sec:model}

\begin{figure}[!ht]
\centering
\includegraphics[width=0.32\textwidth]{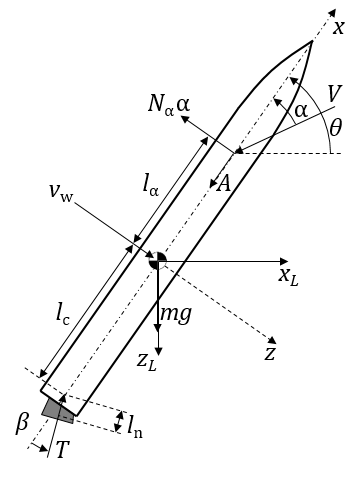}
\caption{Launcher model diagram}
\label{fig:LV_sketch}
\end{figure}

The present study relies on a conventional 3 degrees-of-freedom launcher model in ascent flight featuring lateral drift $z$ and pitch $\theta$ dynamics, as schematised in Fig.~\ref{fig:LV_sketch}. These dynamics, representing the first/second time-derivatives of $z$ as \{$w=\dot{z}$, $\dot{w}=\ddot{z}$\} and the first/second time-derivatives of $\theta$ as \{$q=\dot{\theta}$, $\dot{q}=\ddot{\theta}$\}, are governed by the well-known nonlinear Newton-Euler equations:
\begin{eqnarray}
    m\dot{w}&=&F_{\alpha}+F_\mathrm{c}+F_\mathrm{n}-mg\sin{\theta} \label{eq:forces} \\
    J\dot{q}&=&M_{\alpha}+M_\mathrm{c}+M_\mathrm{n} \label{eq:torques}
\end{eqnarray}
where $m$, $J$ and $g$ are the launcher's mass, lateral moment of inertia and gravity acceleration, \{$F_{\alpha}$, $M_{\alpha}$\} are the aerodynamic force/torque, \{$F_{\mathrm{c}}$, $M_{\mathrm{c}}$\} are the TVC-induced force/torque and \{$F_{\mathrm{n}}$, $M_{\mathrm{n}}$\} are the nozzle moving-mass effects, also known as tail-wags-dog (TWD).

The aerodynamic force and torque are computed as:
\begin{eqnarray}
    F_{\alpha}&=&-SC_{N_{\alpha}}Q\alpha \\
    M_{\alpha}&=&-l_{\alpha}F_{\alpha}
\end{eqnarray}
where $S$, $C_{N_{\alpha}}$ and $l_{\alpha}$ are the reference aerodynamic area, lateral force gradient and aerodynamic arm (distance between the launcher's centres of pressure and gravity). $Q\alpha$ is the aerodynamic load indicator, defined as the product between aerodynamic pressure and angle of attack, which are respectively given by:
\begin{eqnarray}
    Q&=&\dfrac{1}{2}\rho V^2\\
    \alpha&=&\theta+\arctan{\dfrac{w-l_{\alpha}q-v_{\mathrm{w}}}{V}}
    \label{eq:aoa}
\end{eqnarray}
where $\rho$ is the air density, $V$ is the total airspeed and $v_\mathrm{w}$ is the lateral wind turbulence speed. The term $l_{\alpha}q$ is often known as aerodynamic damping.

The TVC-induced force and torque are computed as:
\begin{eqnarray}
    F_\mathrm{c}&=&-T\sin{\beta} \\
    M_\mathrm{c}&=&l_{\mathrm{c}}F_\mathrm{c} 
\end{eqnarray}
where $T$ is the thrust magnitude, $l_{\mathrm{c}}$ is the TVC arm (distance between the launcher's centre of gravity and nozzle's pivot point) and $\beta$ is the TVC deflection angle.

Finally, the nozzle TWD effects are computed as:
\begin{eqnarray}
    F_{\mathrm{n}}&=&-m_{\mathrm{n}}l_{\mathrm{n}}\ddot{\beta} \\
    M_\mathrm{n}&=&l_{\mathrm{c}}F_\mathrm{n}-J_{\mathrm{n}}\ddot{\beta}
\end{eqnarray}
where $m_{\mathrm{n}}$ is the nozzle moving-mass, $l_{\mathrm{n}}$ is the moving-mass arm (distance between the nozzle's centre of gravity and pivot point), $\ddot{\beta}$ is the TVC deflection acceleration and $J_{\mathrm{n}}$ is the nozzle moment of inertia with respect to the pivot point (not to the centre of gravity). 

Most of the model's parameters vary along the launcher's trajectory (this dependence was not evidenced in the previous equations for the sake of readability) and are highly uncertain. These parameters were extracted as a function of time from the simulator presented in~\cite{Pedro_JSR} for a 80~seconds trajectory. The uncertainty levels assumed in this study are summarised in Table~\ref{tab:uncert}.

\begin{table}[!ht]
\centering
\caption{Uncertainty level per type of parameter}
\label{tab:uncert}
\begin{tabular}{|l|c|c|}
\hline
\rowcolor{Gray} Type of parameters & Variables & Uncertainty level \\
\hline
Aerodynamics & $C_{N_{\alpha}}$, $l_{\alpha}$, $\rho$, $V$\ & 20\% \\
Mass/propulsion & $m$, $J$, $l_{\mathrm{c}}$, $T$\ & 10\% \\
\hline
\end{tabular}
\end{table}

Note that while mass/propulsion parameters have an explicit dependency on time, related to the way the propellant burns, aerodynamics parameters have an implicit dependency through intermediate quantities such as altitude and Mach number. 

In addition to the launcher model described above, the present study considers the dynamical effects of TVC actuation and wind turbulence. Both effects are modelled as time-invariant transfer functions for the sake of simplicity without loss of generality. The TVC dynamics corresponds to a second-order system given by:
\begin{equation}
    G_{\mathrm{TVC}}(\mathrm{s})=\dfrac{67.8^2}{\mathrm{s}^2+90.9\,\mathrm{s}+67.8^2}
\end{equation}
where s represents the Laplace variable. The wind turbulence speed $v_\mathrm{w}$ is modelled by colouring a white noise signal through a first-order Dryden filter~\cite{Pedro_AA21} given by:
\begin{equation}
    G_{\mathrm{w}}(\mathrm{s})=\dfrac{3.54}{\mathrm{s}+0.32}
\end{equation}

\begin{figure}[!ht]
\centering
\includegraphics[width=\textwidth]{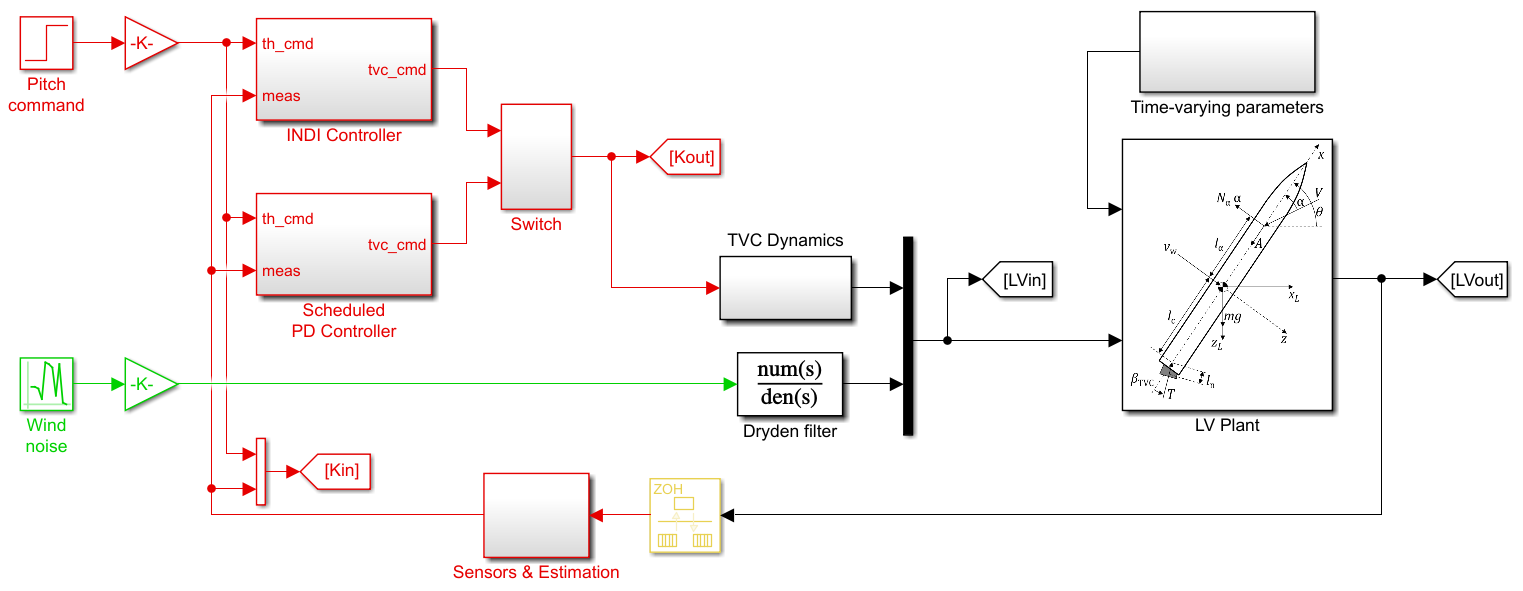}
\caption{Launcher simulator diagram}
\label{fig:simulator}
\end{figure}

The launcher, TVC and wind models were put together in a simulator that allows to quickly analyse and compare several control systems, which is illustrated in Fig.~\ref{fig:simulator}. In this figure, different simulation rates are highlighted using different colours: black for the continuous-time dynamics, red for GNC computations ($f_{\mathrm{GNC}}=25$~Hz, which is well representative of current onboard capabilities) and green for wind noise generation ($f_{\mathrm{w}}=20$~Hz in this case).

For control design purposes, it is also convenient to define a linear model that fully captures the driving dynamics of Eq.~\eqref{eq:forces} and~\eqref{eq:torques}. To do so~\cite{Simplicio2016}, consider the following coefficients relative to the rotational motion:
\begin{equation}
    \mu_{\alpha}=\dfrac{l_\alpha QSC_{N_\alpha}}{J}, \qquad \mu_\mathrm{c}=\dfrac{l_\mathrm{c}T}{J}, \qquad     \mu_\mathrm{n}=\dfrac{m_\mathrm{n}l_\mathrm{n}l_\mathrm{c}+J_\mathrm{n}}{J}
\end{equation}
and to the translational motion:
\begin{equation}
    n_\alpha=\dfrac{QSC_{N_\alpha}}{m}, \qquad n_{\mathrm{c}}=\dfrac{T}{m}, \qquad n_{\mathrm{n}}=\dfrac{m_\mathrm{n}l_\mathrm{n}}{m}
\end{equation}

Using these coefficients, the transfer functions $\beta\mathrm{(s)}\rightarrow\theta\mathrm{(s)}$ and $\beta\mathrm{(s)}\rightarrow w\mathrm{(s)}$ correspond to the solutions of the system:
\begin{equation}
    \begin{bmatrix}
        \mathrm{s}^2+l_{\alpha}\dfrac{\mu_{\alpha}}{V}\mathrm{s}-\mu_{\alpha} & -\dfrac{\mu_{\alpha}}{V} \\
        -l_{\alpha}\dfrac{n_{\alpha}}{V}\mathrm{s}+n_{\alpha}+g\sin{\theta_0} & \mathrm{s}+\dfrac{n_{\alpha}}{V}
    \end{bmatrix}
    \begin{bmatrix}
        \dfrac{\theta (\mathrm{s})}{\beta (\mathrm{s})} \\
        \dfrac{w (\mathrm{s})}{\beta (\mathrm{s})}
    \end{bmatrix}=-
    \begin{bmatrix}
        \mu_\mathrm{n}\mathrm{s}^2+\mu_\mathrm{c} \\
        n_\mathrm{n}\mathrm{s}^2+n_\mathrm{c}
    \end{bmatrix}
\end{equation}

Furthermore, as a first approximation for attitude control design purposes, drift and TWD dynamics can be neglected and the transfer function $\beta\mathrm{(s)}\rightarrow\theta\mathrm{(s)}$ simplifies into:
\begin{equation}
    \dfrac{\theta(\mathrm{s})}{\beta(\mathrm{s})} \approx -\dfrac{\mu_\mathrm{c}}{ \mathrm{s}^2+l_{\alpha}\dfrac{\mu_{\alpha}}{V}\mathrm{s}-\mu_{\alpha}}
    \label{eq:oltf}
\end{equation}

\section{Launcher control design using angular acceleration feedback} \label{sec:design}

This section describes and justifies the four attitude control systems developed in this study. 

\subsection{Scheduled PD controller} \label{sec:K1}

The baseline controller for this study is a classic proportional-derivative (PD) controller with the following structure:
\begin{equation}
    \beta(\mathrm{s})=k_P\Big(\theta_{\mathrm{cmd}}(\mathrm{s})-\theta(\mathrm{s})\Big)-k_D\,q(\mathrm{s})=k_P\,\theta_{\mathrm{cmd}}(\mathrm{s})-\Big(k_P+\mathrm{s}\,k_D\Big)\theta(\mathrm{s})
    \label{eq:pdlaw}
\end{equation}
Despite their simplicity, PD controllers represent the industrial state-of-practice for the vast majority of launch vehicles~\cite{SICE2020}. The gains $k_P$ and $k_D$ can be tuned using a multitude of methods. Here, they are selected based on pole placement of the closed-loop transfer function, which is obtained by substituting Eq.~\eqref{eq:pdlaw} in~\eqref{eq:oltf}:
\begin{equation}
    \dfrac{\theta(\mathrm{s})}{\theta_{\mathrm{cmd}}(\mathrm{s})} = -\dfrac{\mu_\mathrm{c}k_P}{ \mathrm{s}^2+\Big(l_{\alpha}\dfrac{\mu_{\alpha}}{V}-\mu_\mathrm{c}k_D\Big)\mathrm{s}-\Big(\mu_{\alpha}+\mu_\mathrm{c}k_P\Big)}
\end{equation}
It is clear from this equation that $k_P$ and $k_D$ can be chosen so as to enforce the desired natural frequency $\omega_{\theta}$ and damping ratio $\zeta$ (here assumed constant throughout the flight for simplicity without loss of generality). It is also clear that this approach does not allow to specify the steady-state gain (when $\mathrm{s} \rightarrow 0$) independently of the natural frequency as they both depend on $k_P$ only.

In order to handle the wide variation of the model's parameters during the flight, the two gains need to be scheduled throughout the trajectory. To do so, they are pre-computed for a grid of $N=9$ points (spaced every 10~seconds along the trajectory) as:
\begin{equation}
    k_P[i]=-\dfrac{1}{\mu_\mathrm{c}[i]}\Big(\mu_{\alpha}[i]+\omega_{\theta}^2\Big), \qquad k_D[i]=\dfrac{1}{\mu_\mathrm{c}[i]}\Big(l_{\alpha}[i]\dfrac{\mu_{\alpha}[i]}{V[i]}-2 \zeta \omega_{\theta}\Big), \qquad i=1,...,N
\end{equation}
and then linearly interpolated online during the simulation. The robustness of this approach can be increased by scheduling the controller with respect to online measurements/estimates of some of the model's parameters. This is the underlying idea of LPV control~\cite{Diego_LPV}, which is outside the scope of this paper.

\subsection{INDI controller} \label{sec:K2}

In this section, an INDI-based control law is developed and applied to regulate the launcher's attitude channel, i.e. to:
\begin{equation}
    y=h(\xr)=q
\end{equation}
where $\xr$ represents the state vector. In order to apply the INDI technique, this equation has to be time-differentiated until an explicit dependency on the TVC input appears. The first-order derivative corresponds to Eq.~\eqref{eq:torques}, which can be recast as:
\begin{equation}
    \dot{y}=\dot{q}=f(\xr)+g(\xr)u
\end{equation}
where $f(\xr)$ is the control-independent part of the model, $g(\xr)$ expresses the influence of the controls in the system and $u$ is the control input. For the launcher scenario, the latter two terms correspond to:
\begin{equation}
     g(\xr)\approx-\mu_\mathrm{c}, \qquad u=\beta
\end{equation}

A virtual control input can now be defined in order to transform the nonlinear system into a linear form as follows:
\begin{equation}
    \nu=\dot{q}=\ddot{\theta} \quad \Rightarrow \quad \dfrac{\theta(\mathrm{s})}{\nu(\mathrm{s})}=\dfrac{1}{\mathrm{s}^2} \label{eq:linearised}
\end{equation}
Following the procedure of Sec.~\ref{sec:intro_indi}, the command signal sent to the TVC actuator is given by:
\begin{equation}
    \beta=\beta_0-\dfrac{1}{\mu_\mathrm{c}}\Big(\nu-\dot{q}_0\Big) \label{eq:linearisation}
\end{equation}
where $\beta_0$ and $\dot{q}_0$ are measurements/estimates of the TVC command and angular acceleration at the current computation step, respectively. The estimate of the TVC command $\beta_0$ is obtained with a low pass filter and because angular acceleration sensors are not common in launchers today, $\dot{q}_0$ is estimated by passing the angular rate $q$ through a derivative filter of the form:
\begin{equation}
    H_{\dot{q}}(\mathrm{s})=\dfrac{\mathrm{s}\,\omega_{\dot{q}}}{\mathrm{s}+\omega_{\dot{q}}} \label{eq:derivfilt}
\end{equation}
where $\omega_{\dot{q}}$ represents the filter bandwidth. Note that, after the feedback linearisation of Eq.~\eqref{eq:linearisation}, there are still some degrees of internal dynamics in the system related to the drift motion and TWD effect, but these dynamics are known to be stable and can be further handled by outer control loops.

\begin{figure}[!ht]
\centering
\includegraphics[scale=0.5]{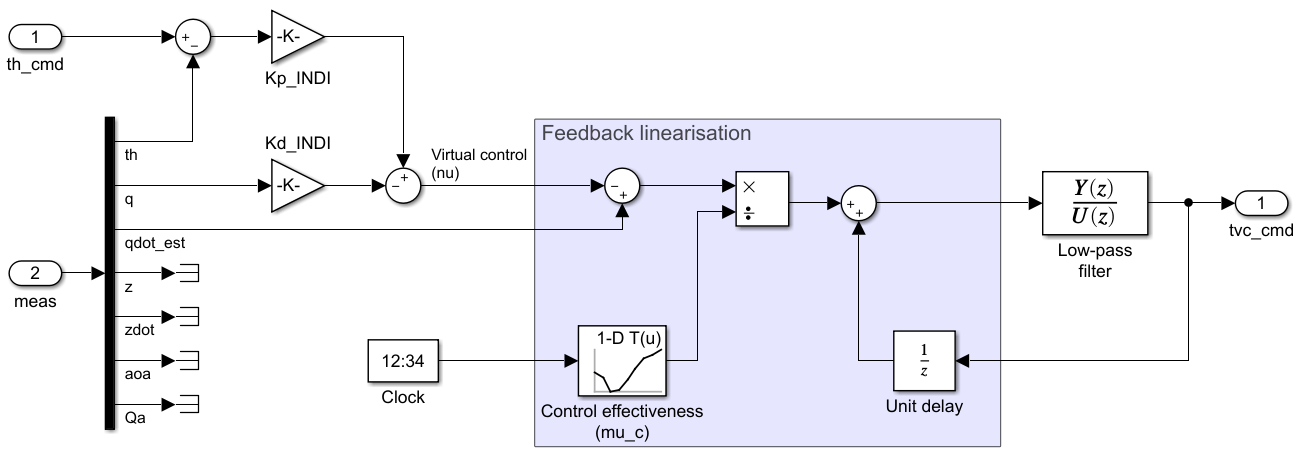}
\caption{INDI controller implementation diagram}
\label{fig:INDI_implement}
\end{figure}

Using the virtual control and the linearised system of Eq.~\eqref{eq:linearised}, an outer PD control law is able to enforce the desired closed-loop response as follows:
\begin{equation}
    \nu(\mathrm{s})=k_P\Big(\theta_{\mathrm{cmd}}(\mathrm{s})-\theta(\mathrm{s})\Big)-k_D\,q(\mathrm{s}) \quad \Rightarrow \quad \dfrac{\theta(\mathrm{s})}{\theta_{\mathrm{cmd}}(\mathrm{s})} = \dfrac{k_P}{ \mathrm{s}^2+k_D\,\mathrm{s}+k_P} \label{eq:indigains}
\end{equation}
\begin{equation}
    k_P=\omega^2_\theta, \qquad k_D=2\zeta\omega_\theta
\end{equation}
Note that, in contrast with the PD controller of Sec.~\ref{sec:K1}, $k_P$ and $k_D$ do not need to be scheduled as they depend on $\omega_\theta$ and $\zeta$ only, but a pre-computed grid of $\mu_\mathrm{c}[i]$ is still required to perform the feedback linearisation. This is highlighted in the blue area of Fig.~\ref{fig:INDI_implement}, which illustrates the implementation of the INDI controller in the simulator. Alternatively, $\mu_\mathrm{c}$ could be estimated based on online measurements. 

\subsection{Scheduled PD controller with $\dot{q}$ feedback} \label{sec:K3}

As explained in Sec.~\ref{sec:intro_indi}, the INDI controller of Eq.~\eqref{eq:linearisation} relies on $\dot{q}$ information to reduce the impact of the launcher's model on the achievable control performance. For a fair comparison of controllers, it is then pertinent to consider a linear controller where $\dot{q}$ feedback is also employed. In this case, the control law takes the form:
\begin{equation}
     \beta(\mathrm{s})=k_P\Big(\theta_{\mathrm{cmd}}(\mathrm{s})-\theta(\mathrm{s})\Big)-k_D\,q(\mathrm{s})-k_A\,\dot{q}(\mathrm{s})
    \label{eq:pdalaw}
\end{equation}
where $k_A$ is the acceleration feedback gain. Similar to Sec.~\ref{sec:K1}, the three gains can be tuned via pole placement of the closed-loop transfer function, which is obtained by substituting Eq.~\eqref{eq:pdalaw} in~\eqref{eq:oltf}:
\begin{equation}
    \dfrac{\theta(\mathrm{s})}{\theta_{\mathrm{cmd}}(\mathrm{s})} = -\dfrac{k_P}{1-\mu_\mathrm{c}k_A}\,\dfrac{\mu_\mathrm{c}}{ \mathrm{s}^2+\dfrac{l_{\alpha}\frac{\mu_{\alpha}}{V}-\mu_\mathrm{c}k_D}{1-\mu_\mathrm{c}k_A}\mathrm{s}-\dfrac{\mu_{\alpha}+\mu_\mathrm{c}k_P}{1-\mu_\mathrm{c}k_A}}
\end{equation}
In contrast with the pure PD controller, the $\dot{q}$ feedback allows to minimise tracking errors because the desired steady-state gain $G_0$ be specified independently of $\omega_\theta$ through the proportional gain as follows:
\begin{equation}
    k_P[i]=\dfrac{\mu_\alpha[i]}{\mu_{\mathrm{c}}[i]}\dfrac{G_0}{1-G_0}, \qquad i=1,...,N
\end{equation}
which is scheduled along a grid of $N=9$ points along the launcher's trajectory. The other two gains are then derived as a function of $\omega_\theta$ and $\zeta$ as:
\begin{equation}
    k_A[i]=\dfrac{1}{\mu_{\mathrm{c}}[i]}\Big(1+\dfrac{\mu_{\alpha}[i]+\mu_\mathrm{c}[i]k_P[i]}{\omega_\theta^2}\Big), \quad k_D[i]=\dfrac{1}{\mu_\mathrm{c}[i]}\Big(l_{\alpha}[i]\dfrac{\mu_{\alpha}[i]}{V[i]}-2 \zeta \omega_{\theta}\left( 1-\mu_\mathrm{c}[i]k_A[i] \right)\Big)
\end{equation}

\begin{figure}[!ht]
\centering
\includegraphics[width=0.5\textwidth]{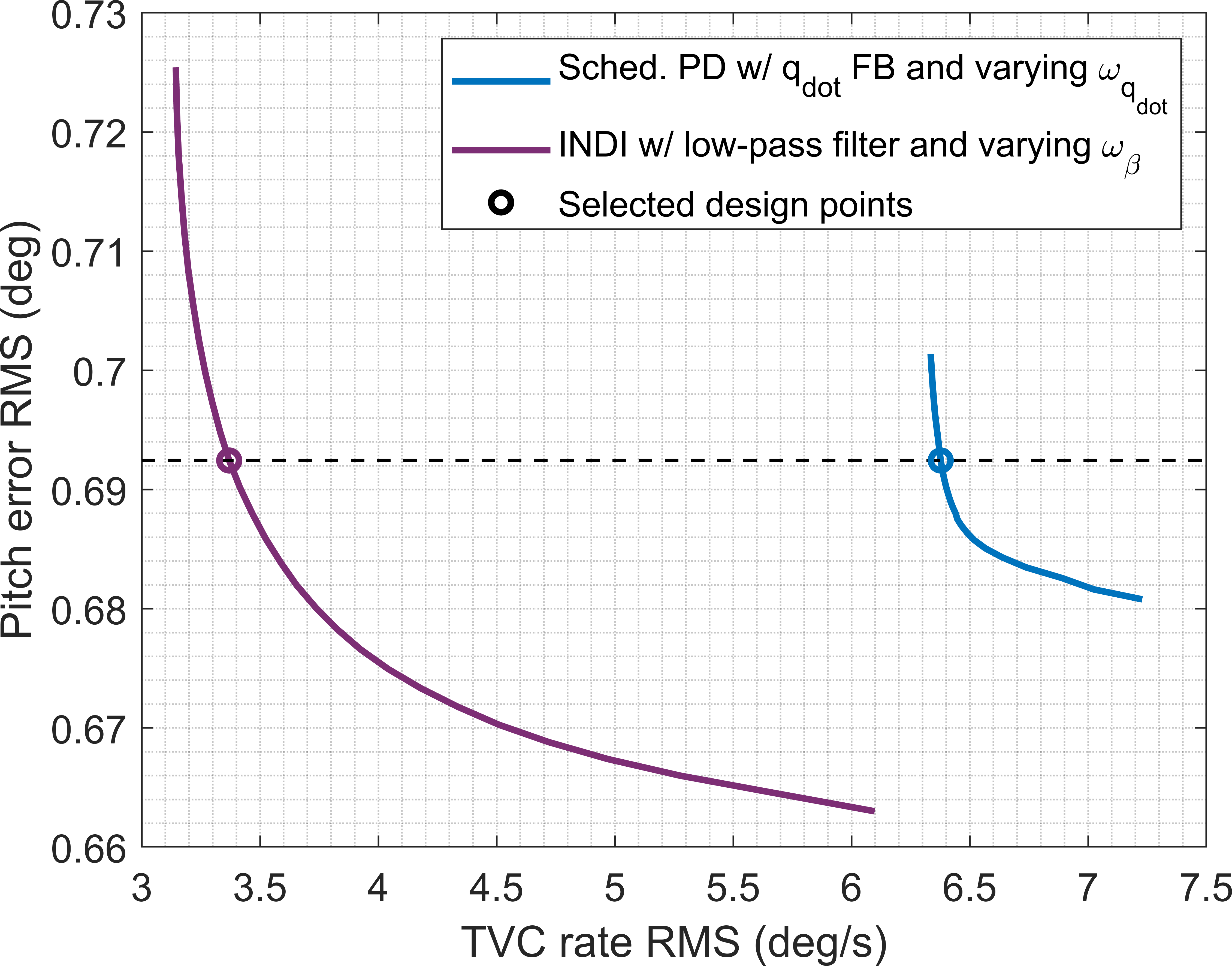}
\caption{Tuning trade-off of angular acceleration feedback approaches}
\label{fig:pareto}
\end{figure}

For the estimation of $\dot{q}$ in Eq.~\eqref{eq:pdalaw}, the same approach of Sec.~\ref{sec:K2}, i.e. passing the angular rate $q$ through the first-order derivative filter of Eq.~\eqref{eq:derivfilt}, was followed. In practice, it was verified that the performance of this controller is fairly sensitive to the filter bandwidth $\omega_{\dot{q}}$. This impact is illustrated in Fig.~\ref{fig:pareto}, which shows root-mean-square (RMS) values of pitch error ($\theta_{\mathrm{err}}=\theta_{\mathrm{cmd}}-\theta$) vs. TVC rate ($\dot{\beta}$) for a step command in $\theta_{\mathrm{cmd}}$ using different controllers and nominal conditions.

The blue line in Fig.~\ref{fig:pareto} shows results using the scheduled PD controller with $\dot{q}$ feedback (FB) and varying values of the derivative filter bandwidth $\omega_{\dot{q}}$. Based on the results, the selection of $\omega_{\dot{q}}$ provides a key (and intuitive) tuning trade-off: increasing the bandwidth leads to smaller errors at the expense of more demanding TVC actuation, and vice-versa. A more favourable trade-off would likely be achieved by using a higher-order derivative filter, which is outside the scope of this paper.

\subsection{INDI controller with low-pass filter} \label{sec:K4}

When applied to the pure INDI controller developed in Sec.~\ref{sec:K2}, the same tuning trade-off analysis showed a much smaller sensitivity to $\omega_{\dot{q}}$ but unacceptably high TVC rates. To address this issue, the INDI controller was augmented with a low-pass filter at the output of the feedback linearisation loop, as depicted on the right-hand side of Fig.~\ref{fig:INDI_implement}.

The feedback linearisation loop, outer linear gains and $\dot{q}$ estimation filter remain unchanged. The low-pass filter has bandwidth $\omega_{\beta}$ and a first-order structure as follows:
\begin{equation}
    H_{\beta}(\mathrm{s})=\dfrac{\omega_{\beta}}{\mathrm{s}+\omega_{\beta}}
\end{equation}

The purple line in Fig.~\ref{fig:pareto} shows the tuning trade-off using the INDI controller with low-pass filter and varying values of its bandwidth $\omega_{\beta}$. Comparing with the PD controller with $\dot{q}$ feedback (blue line), the two controllers show a similar trend (i.e. smaller errors and larger TVC rates for higher bandwidths), yet the INDI controller leads to smaller TVC rates for the same level of error. As before, a more favourable trade-off would likely be achieved by using a higher-order low-pass filter, but this is outside the scope of the paper.

\subsection{Control design summary}

The four controllers in Sec.~\ref{sec:K1} to~\ref{sec:K4} have been designed so as to enforce the same closed-loop properties throughout the flight. These are:
\begin{itemize}
    \item Natural frequency $\omega_\theta=2.5$~rad/s;
    \item Damping ratio $\zeta=0.8$;
    \item Steady-state error of 5\%, i.e. $G_0=1.05$, only applicable to Sec.~\ref{sec:K3}.
\end{itemize}
Furthermore, the bandwidth of the filters in Sec.~\ref{sec:K3} and~\ref{sec:K4}, $\omega_{\dot{q}}$ and $\omega_\beta$, has been tuned so as to provide the same pitch error in nominal conditions, as highlighted in Fig.~\ref{fig:pareto}. The robust performance of these controllers will then be analysed in Sec.~\ref{sec:perf}.

Table~\ref{tab:depend} provides an overview of each controller's dependency on the model parameters and sensor measurements/estimates. As anticipated, from the scheduled PD controller to the INDI controller, there is a progressive reduction of model dependency and increased use of sensor information. More specifically, the INDI controller relies on measurements/estimates of $\dot{q}$ and $\beta$ to fully circumvent the knowledge of the aerodynamics model. 

\begin{table}[!ht]
\centering
\caption{Dependencies per control design method}
\label{tab:depend}
\begin{tabular}{|l|l|l|}
\hline
\rowcolor{Gray} Control design & Dependency on & Dependency on \\
\rowcolor{Gray} method & model parameters & measurements/estimates \\
\hline
Scheduled PD & $J,\,l_{\mathrm{c}},\,T,\,C_{N_{\alpha}},\,l_{\alpha},\,\rho,\,V$ & $\theta,\,q$ \\
Scheduled PD with $\dot{q}$ feedback & $J,\,l_{\mathrm{c}},\,T,\,C_{N_{\alpha}},\,l_{\alpha},\,\rho,\,V$ & $\theta,\,q,\,\dot{q}$ \\
INDI with or without low-pass filter & $J,\,l_{\mathrm{c}},\,T$ & $\theta,\,q,\,\dot{q},\beta$ \\
\hline
\end{tabular}
\end{table}

\section{Time-domain robust performance analysis} \label{sec:perf}

\begin{figure}[!p]
\centering
{\includegraphics[width=0.32\textwidth]{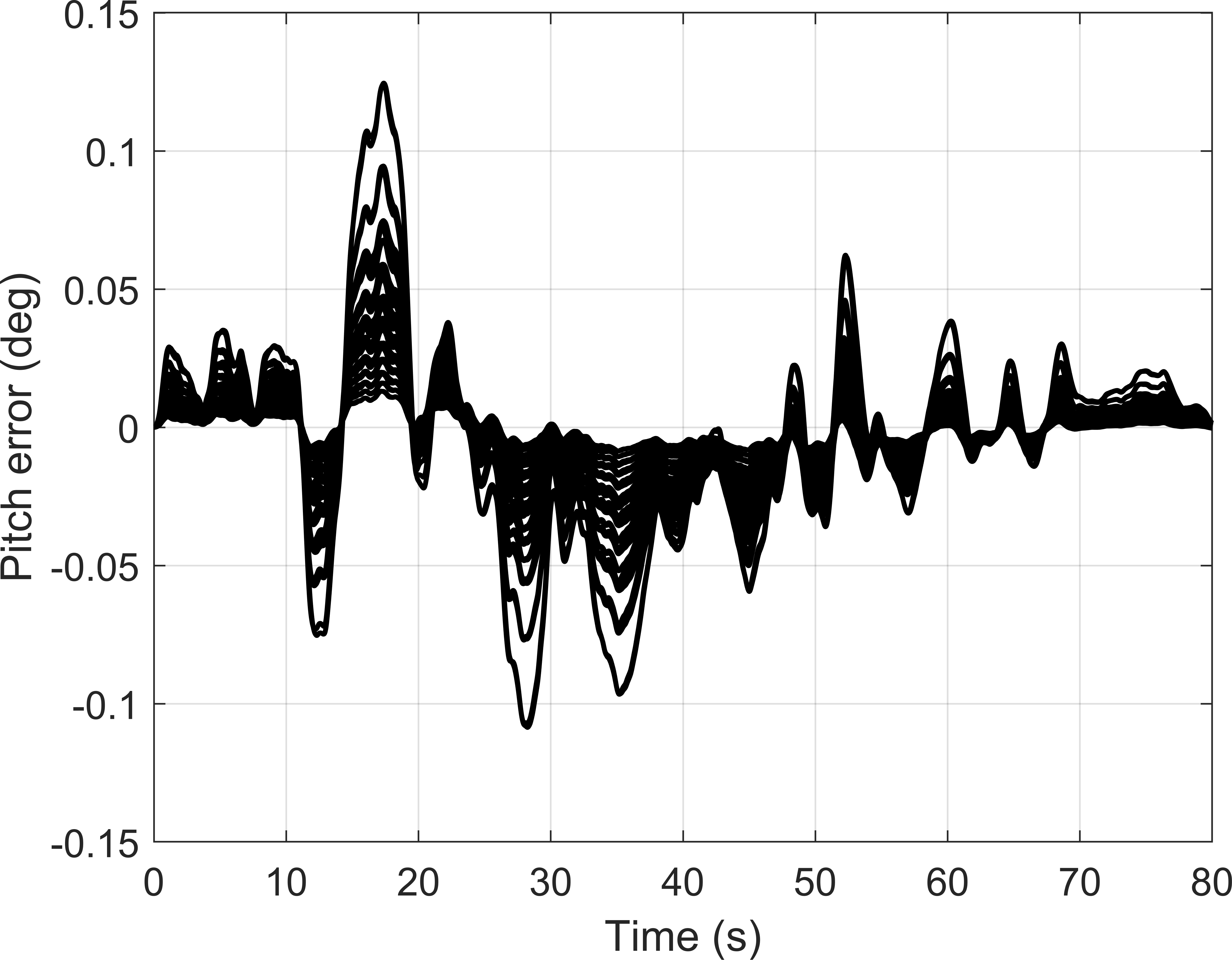}} \hfill
{\includegraphics[width=0.32\textwidth]{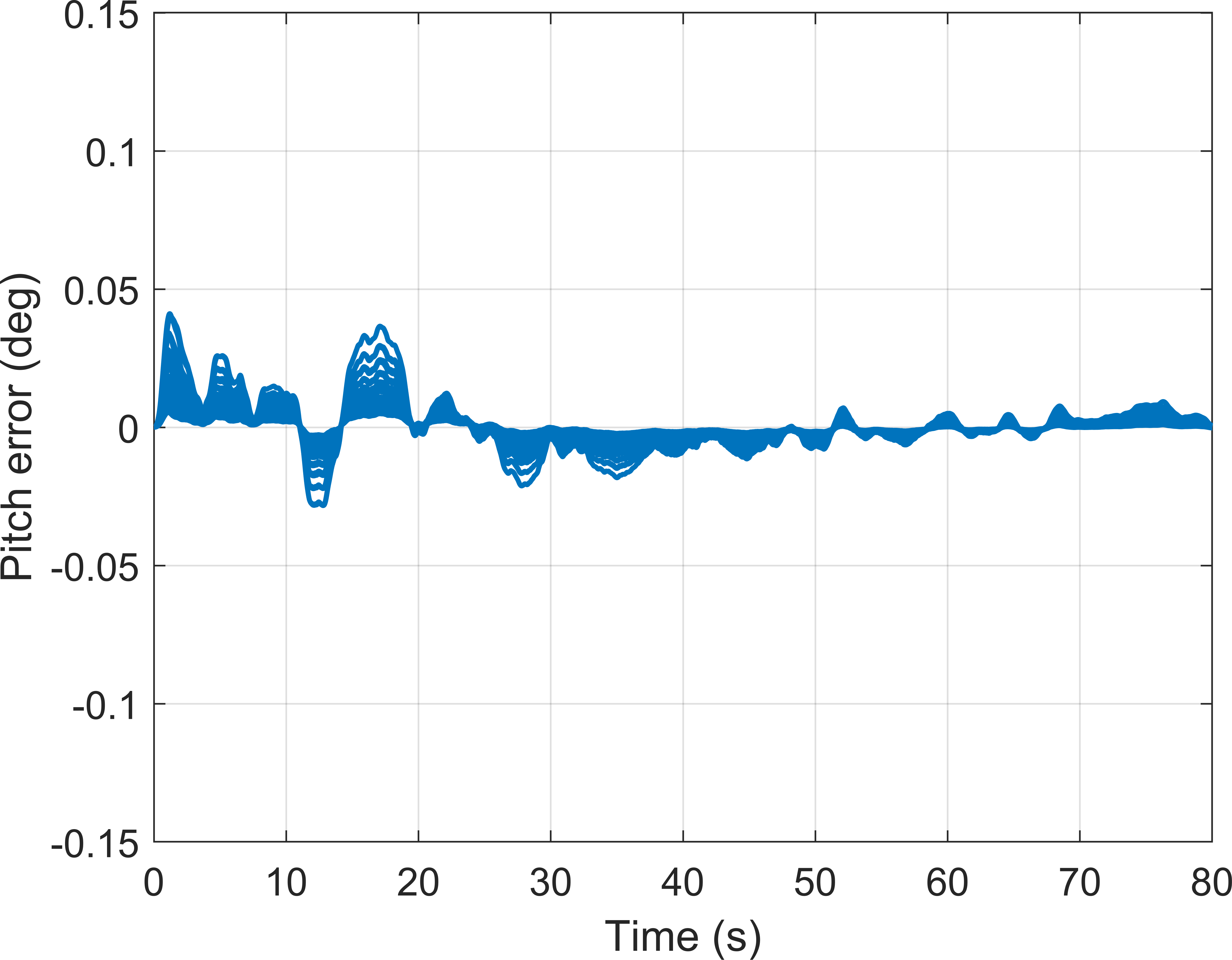}} \hfill
{\includegraphics[width=0.32\textwidth]{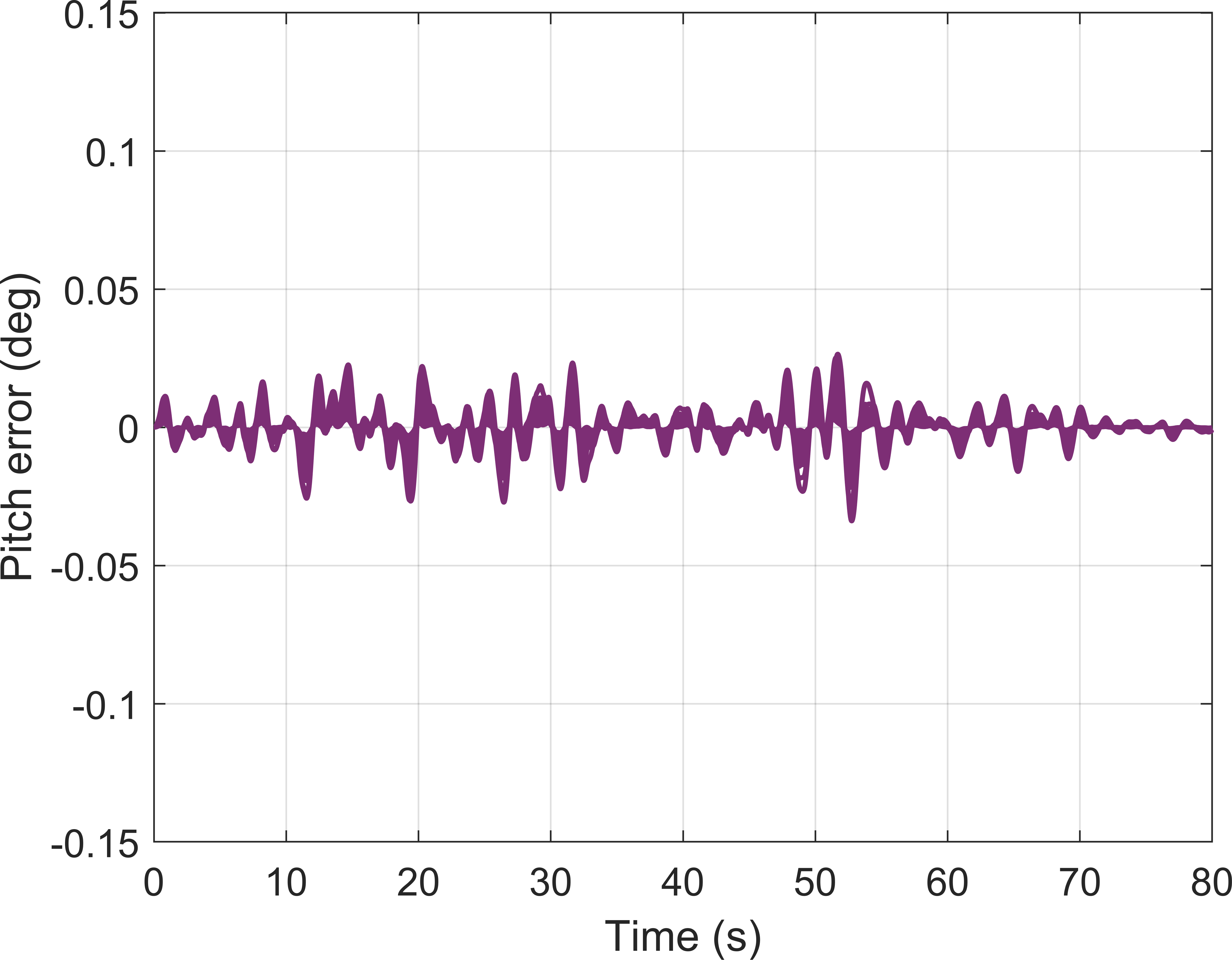}} \\ \vspace{1mm}
{\includegraphics[width=0.32\textwidth]{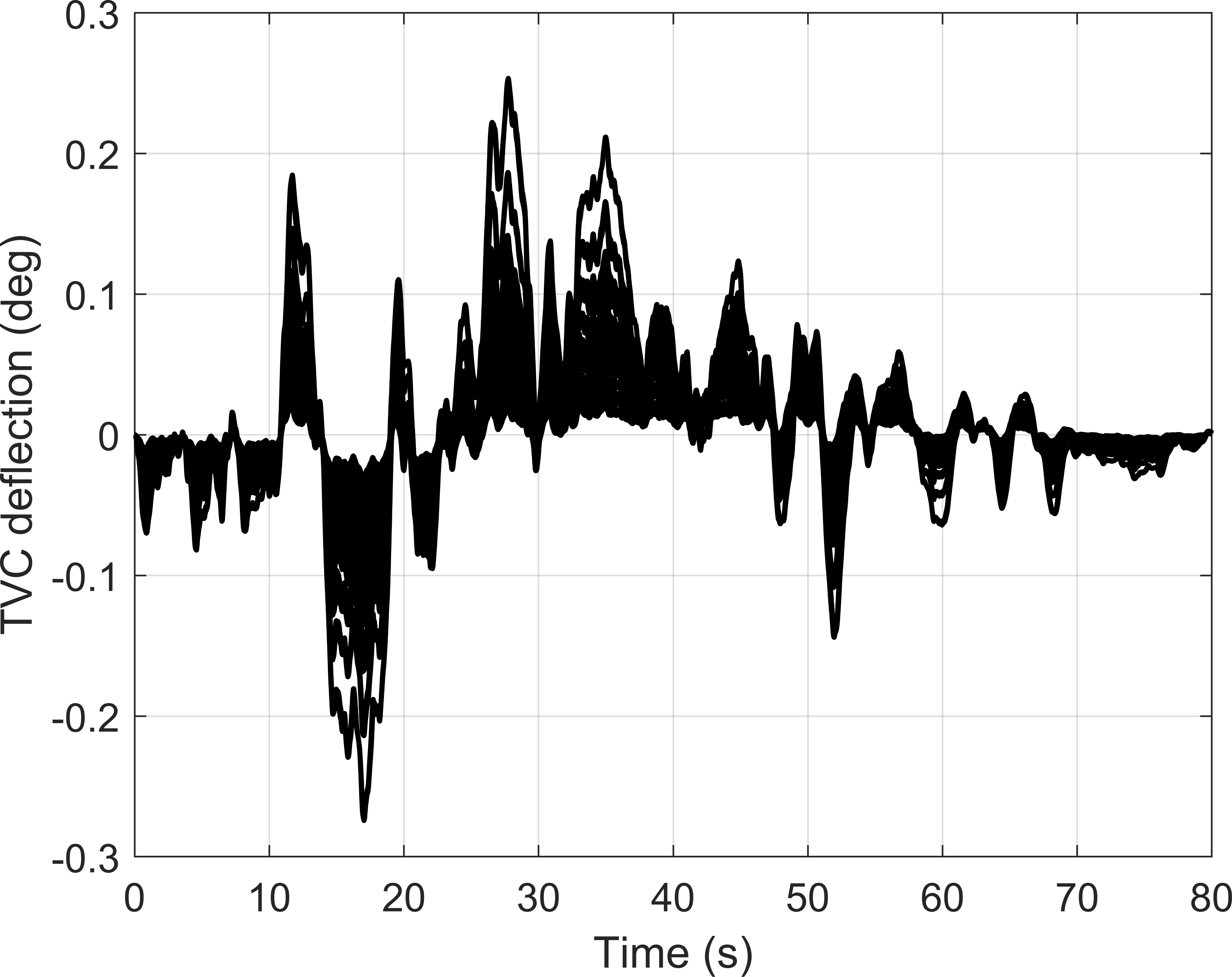}} \hfill
{\includegraphics[width=0.32\textwidth]{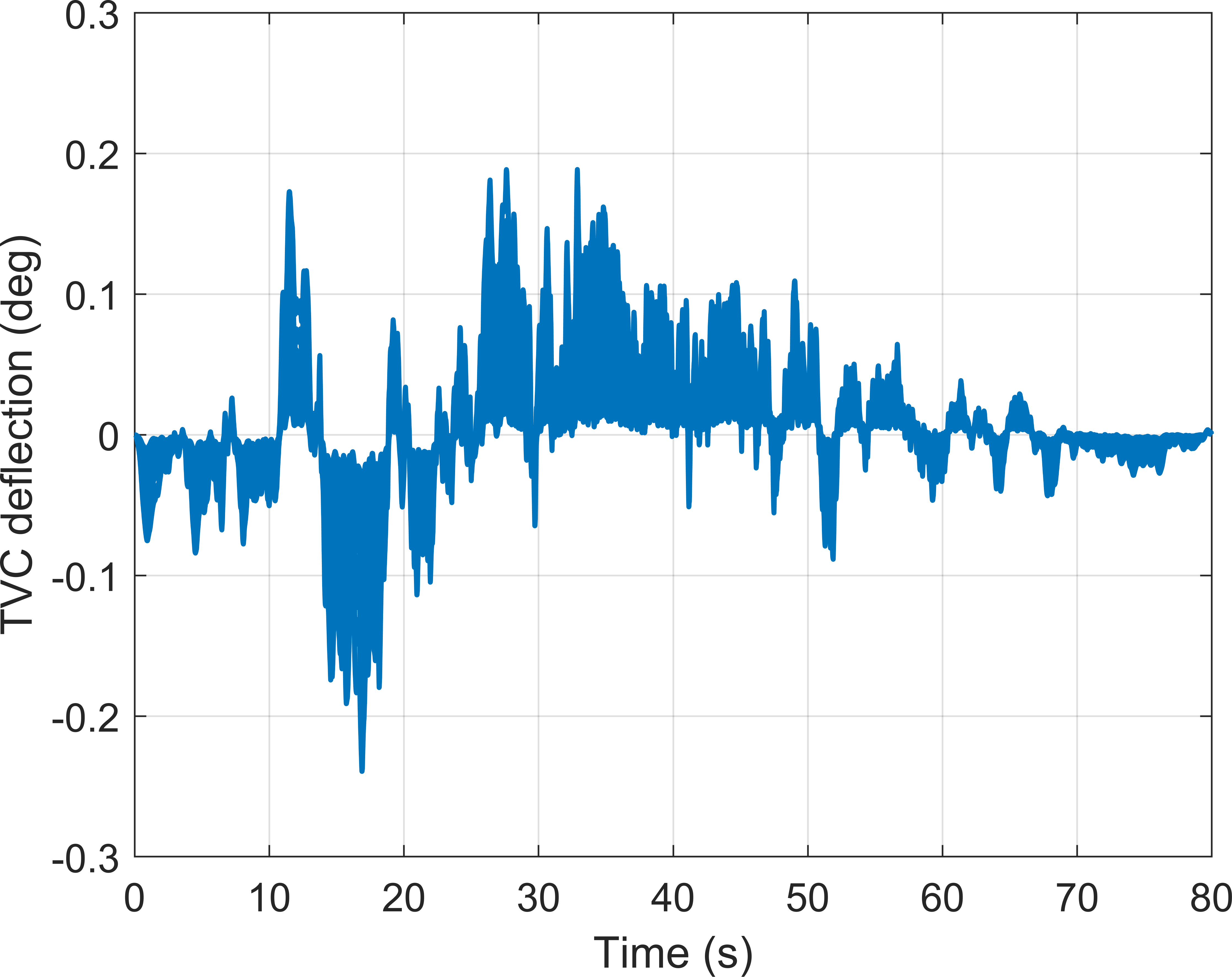}} \hfill
{\includegraphics[width=0.32\textwidth]{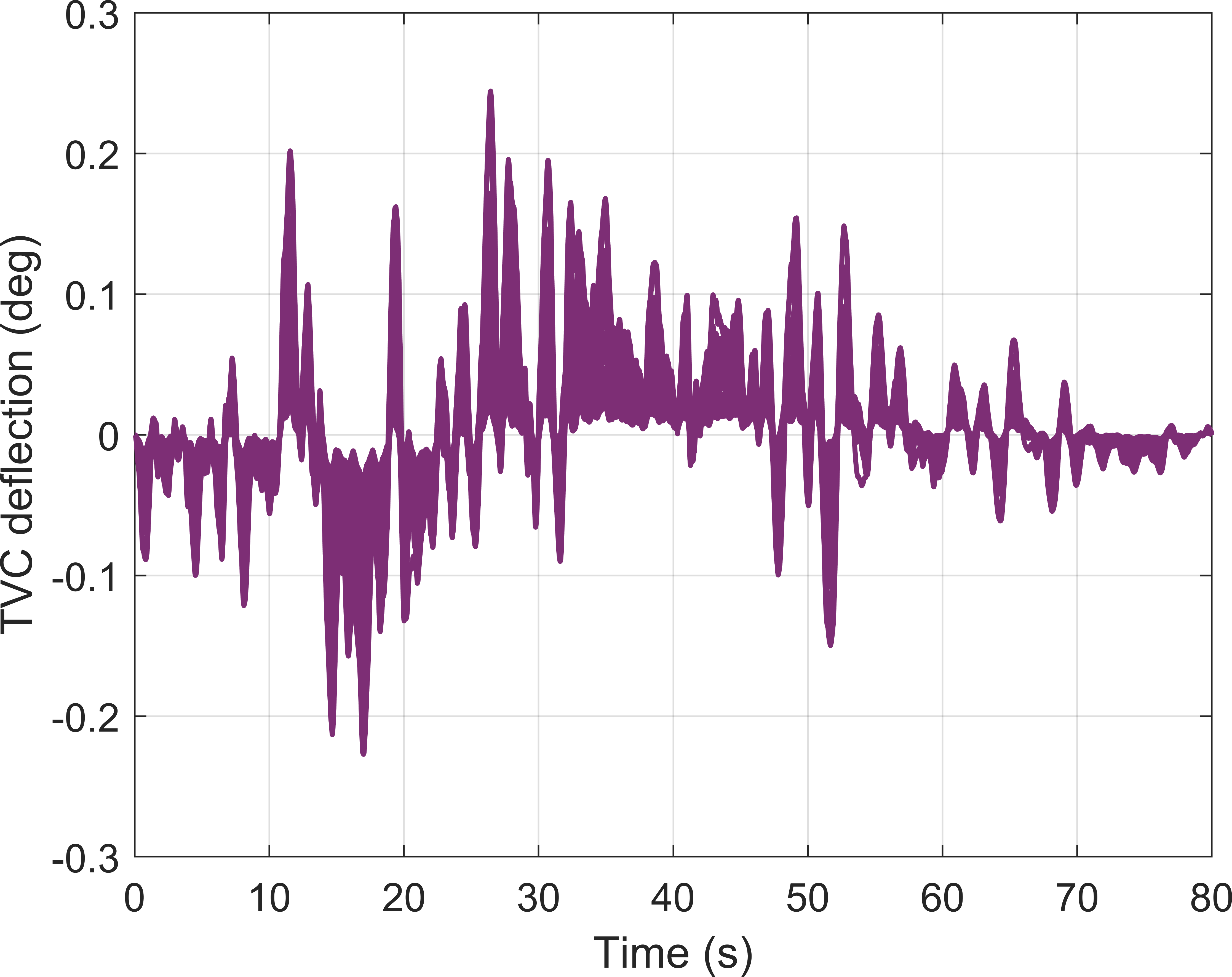}} \\ \vspace{-4mm}
\subfloat[Scheduled PD controller]{\includegraphics[width=0.32\textwidth]{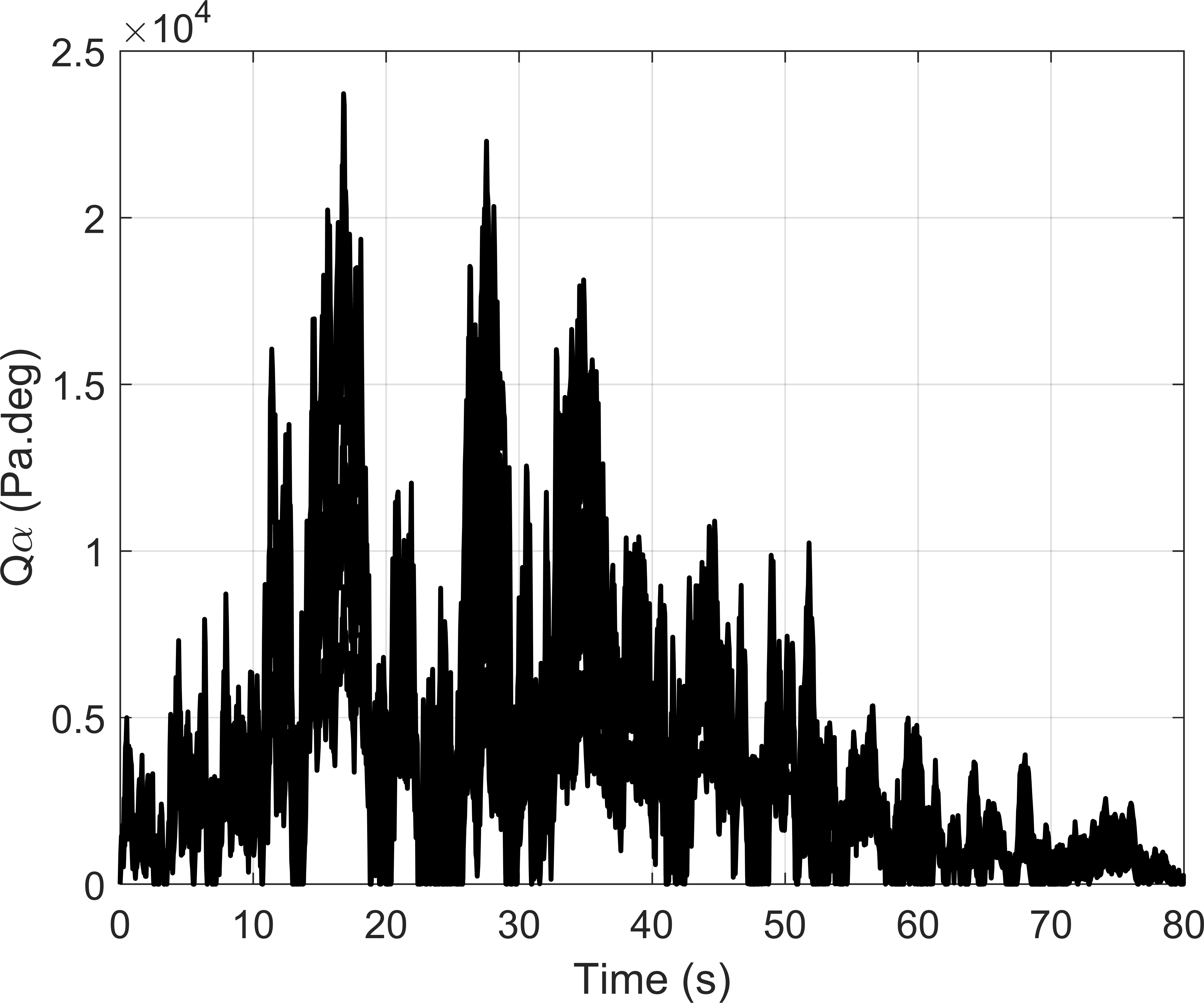}} \hfill
\subfloat[Scheduled PD controller w/ $\dot{q}$ FB]{\includegraphics[width=0.32\textwidth]{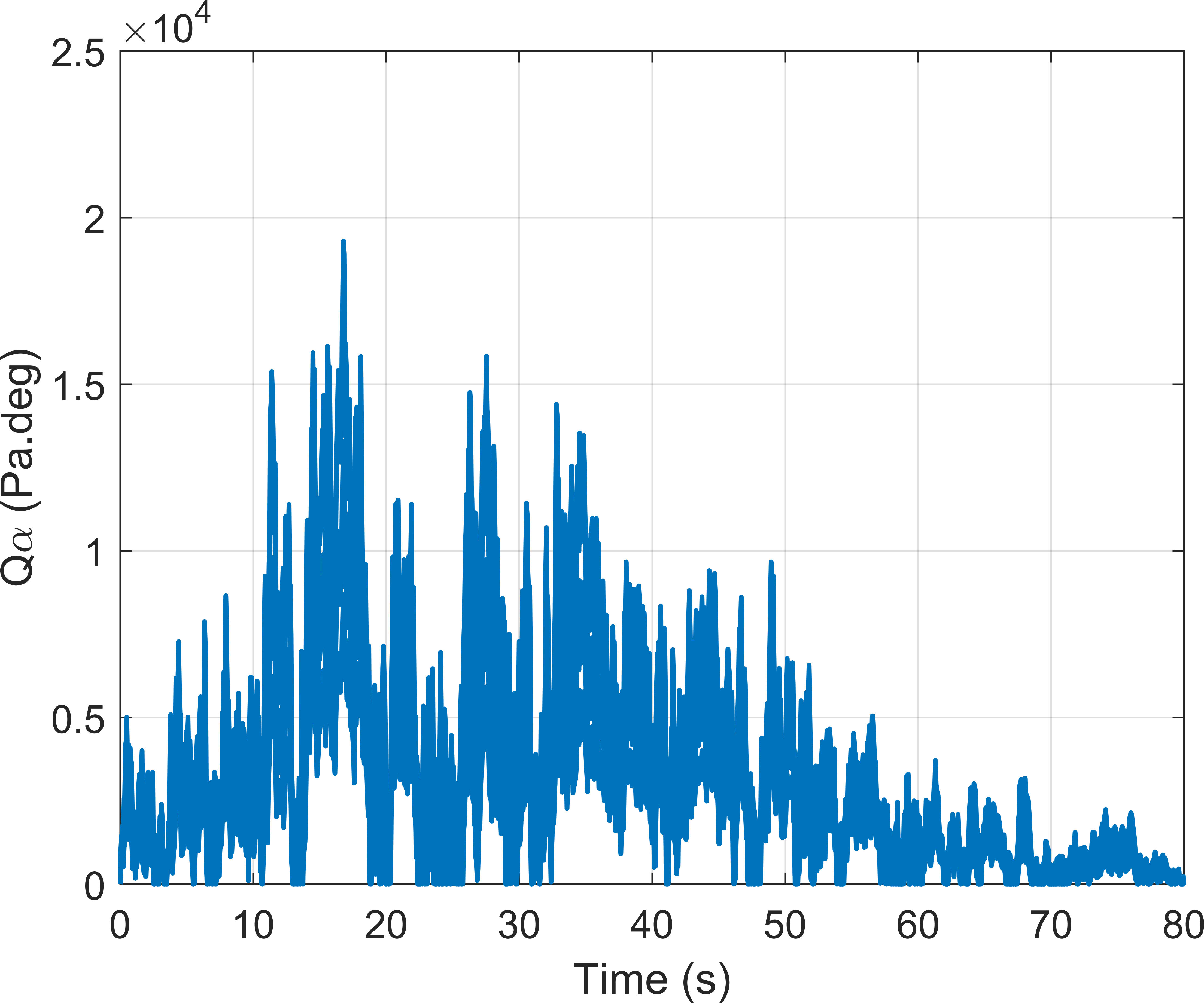}} \hfill
\subfloat[INDI controller w/ low-pass filter]{\includegraphics[width=0.32\textwidth]{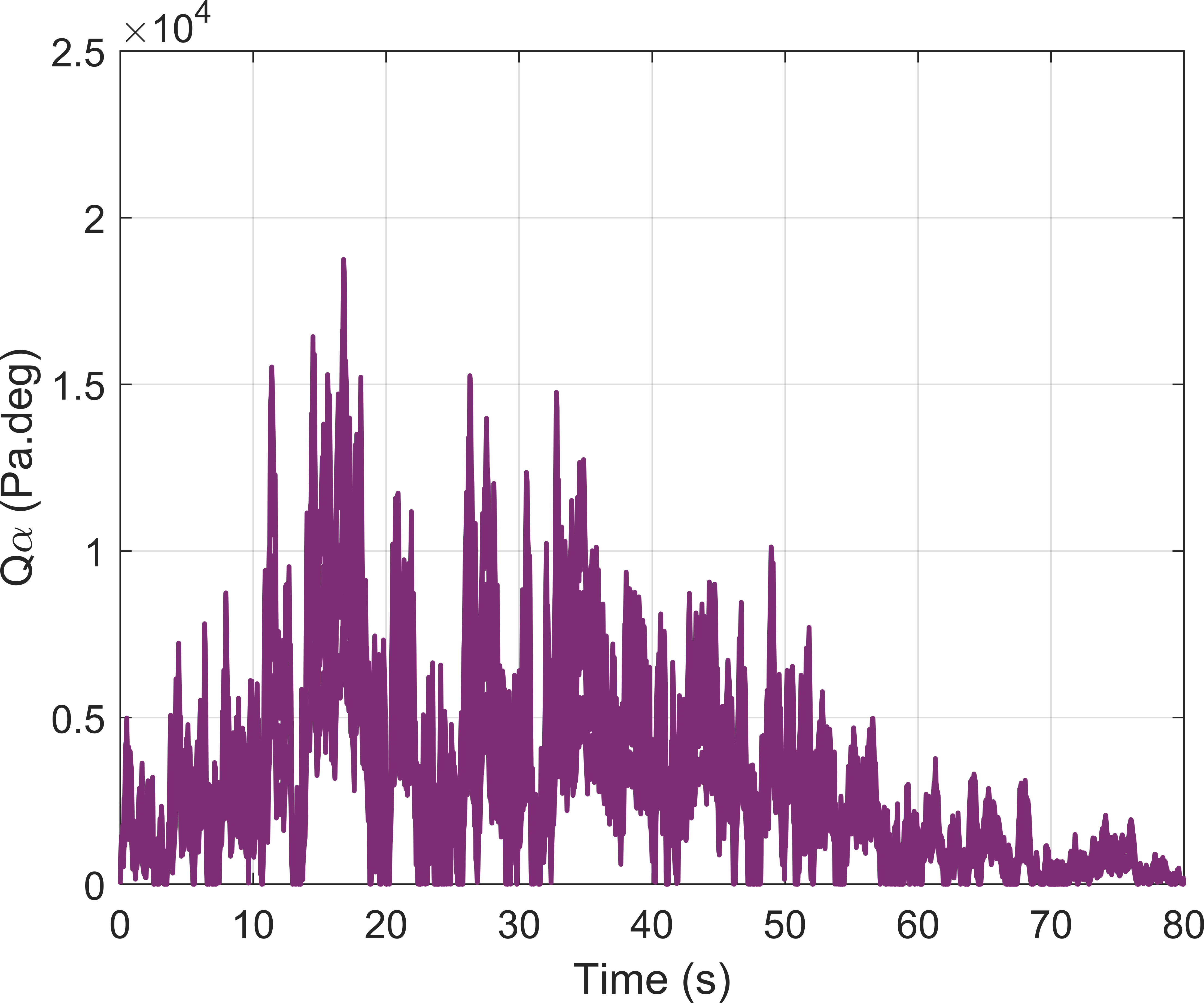}} 
\caption{Comparison of Monte-Carlo wind responses using different controllers}
\label{fig:sims}
\end{figure}

This section analyses and compares the nonlinear time-domain performance the controllers developed in Sec.~\ref{sec:design}. Figure~\ref{fig:sims} shows dispersed responses of the $2^8=256$ corner-cases within the uncertainty level of Table~\ref{tab:uncert} when subjected to the same wind turbulence input $v_\mathrm{w}$, modelled as described in Sec.~\ref{sec:model}.

From the top to the bottom rows, the figure depicts the obtained pitch error $\theta_{\mathrm{err}}$, TVC deflection $\beta$ and aerodynamic load indicator $Q\alpha$ along the trajectory. From left to right, the figure depicts results using the scheduled PD controller (Fig.~\ref{fig:sims}a, in black), scheduled PD controller with $\dot{q}$ feedback (Fig.~\ref{fig:sims}b, in blue) and INDI controller with low-pass filter (Fig.~\ref{fig:sims}c, in purple). The pure INDI controller (wihout low-pass filter) is not shown as it leads to unacceptably high TVC rates.

From Fig.~\ref{fig:sims}a to~\ref{fig:sims}b, a reduction in the dispersion of all the indicators can be observed. This joint reduction clearly demonstrates the benefit of including $\dot{q}$ feedback in the control design. The pitch error (and partially the $Q\alpha$) is further reduced when using the INDI controller with low-pass filter, as depicted in Fig.~\ref{fig:sims}c, at the expensive of higher TVC deflections (although still comparable to the pure PD controller). Note that $Q\alpha$ minimisation was not a specific control design objective in this case, but comes as a direct consequence of smaller pitch and drift errors, as indicated in Eq.~\eqref{eq:aoa}.

\begin{figure}[!p]
\centering
\subfloat[Wind responses]
{\includegraphics[width=0.49\textwidth]{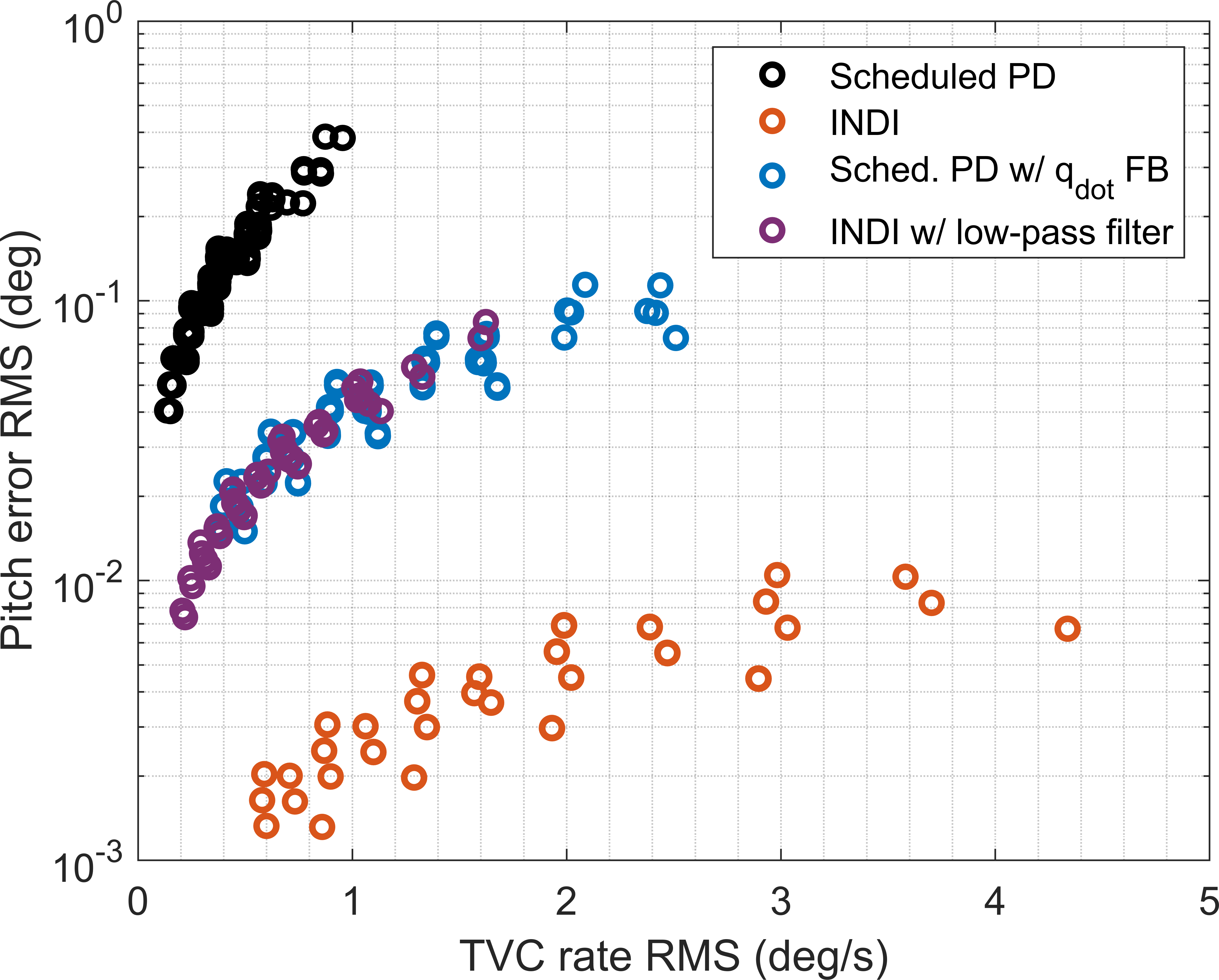}} \hfill
\subfloat[Step responses]
{\includegraphics[width=0.49\textwidth]{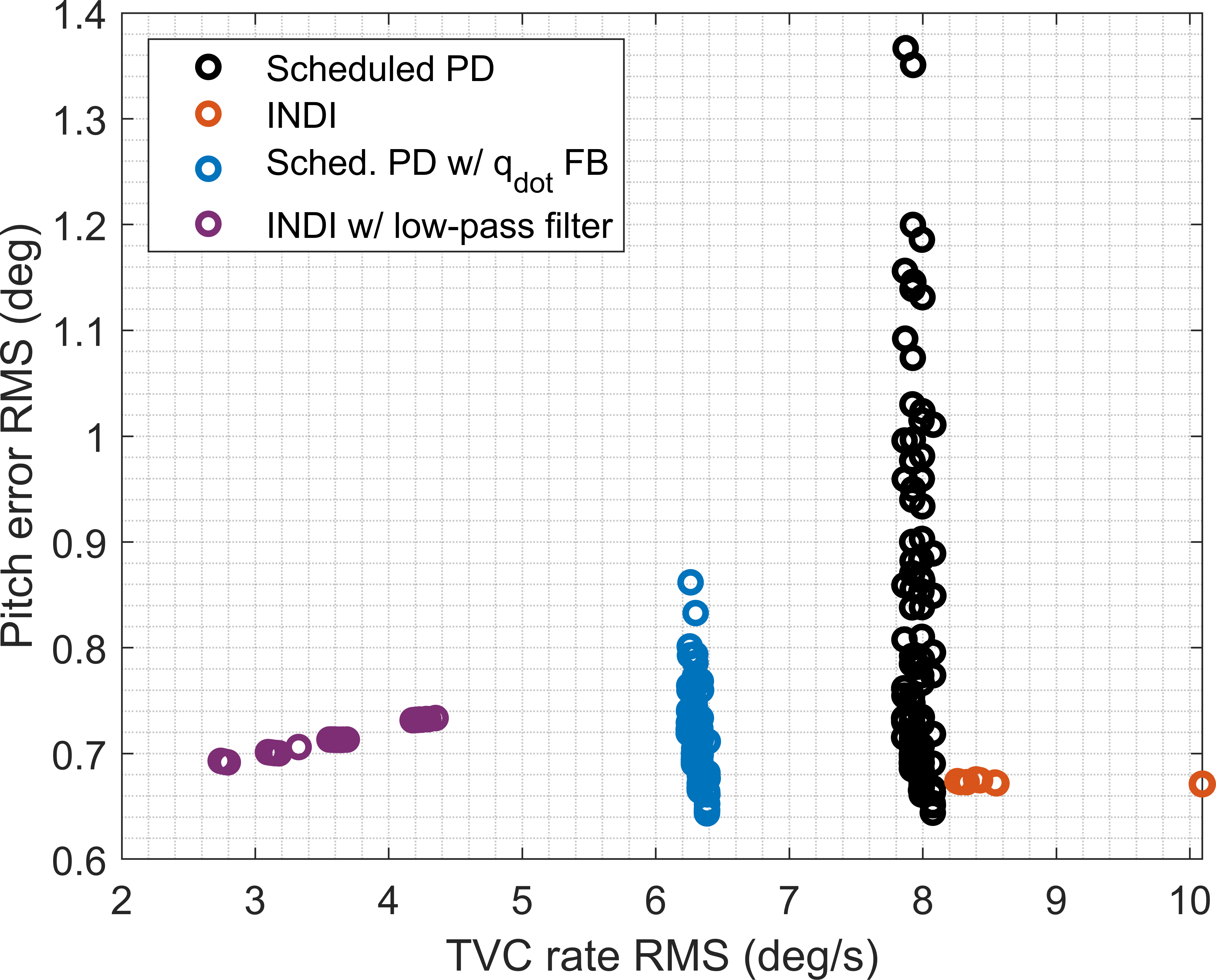}}
\caption{Overview of Monte-Carlo results using different controllers}
\label{fig:results}
\end{figure}

In order to more clearly visualise these trends, Fig.~\ref{fig:results}a shows the wind response results using the same RMS $\theta_{\mathrm{err}}$ vs. $\dot{\beta}$ plot of Fig.~\ref{fig:pareto}. Each point in Fig.~\ref{fig:results}a corresponds to a single simulation from Fig.~\ref{fig:sims}.

As anticipated, the pure PD controller (in black) provides the largest errors but the smallest TVC rates while, on the other hand, the pure INDI controller (in red) provides the smallest errors but the largest TVC rates. The PD controller with $\dot{q}$ feedback (in blue) and the INDI controller with low-pass filter (in purple) lie in-between the two extremes, with the latter controller performing better than the former (i.e. with slightly smaller errors and TVC rates) but only marginally.

In order to complement the analysis, Fig.~\ref{fig:results}b shows the same type of results for a step command in $\theta_\mathrm{cmd}$. As before, the pure PD controller (in black) leads by far to the largest errors and the pure INDI controller (in red) to the largest TVC rates. Performance in terms of error and TVC rate improves using either the PD controller with $\dot{q}$ feedback or the INDI controller with low-pass filter, and the difference between these two controllers is now more evident than for the wind responses.

In nominal conditions, it was known from Fig.~\ref{fig:pareto} that, for the same error, the INDI controller with low-pass filter (in purple) provides a smaller TVC rate than the PD controller with $\dot{q}$ feedback (in blue). Nonetheless, Fig.~\ref{fig:results}a shows that the former controller performs better also in terms of error, having a range of dispersion that is approximately four times smaller. The smaller error dispersion of the INDI controller with low-pass filter comes at the expense of a larger TVC rate dispersion, but its maximum value remains significantly lower than that of the PD controller with $\dot{q}$ feedback.

\begin{figure}[!ht]
\centering
\includegraphics[width=0.5\textwidth]{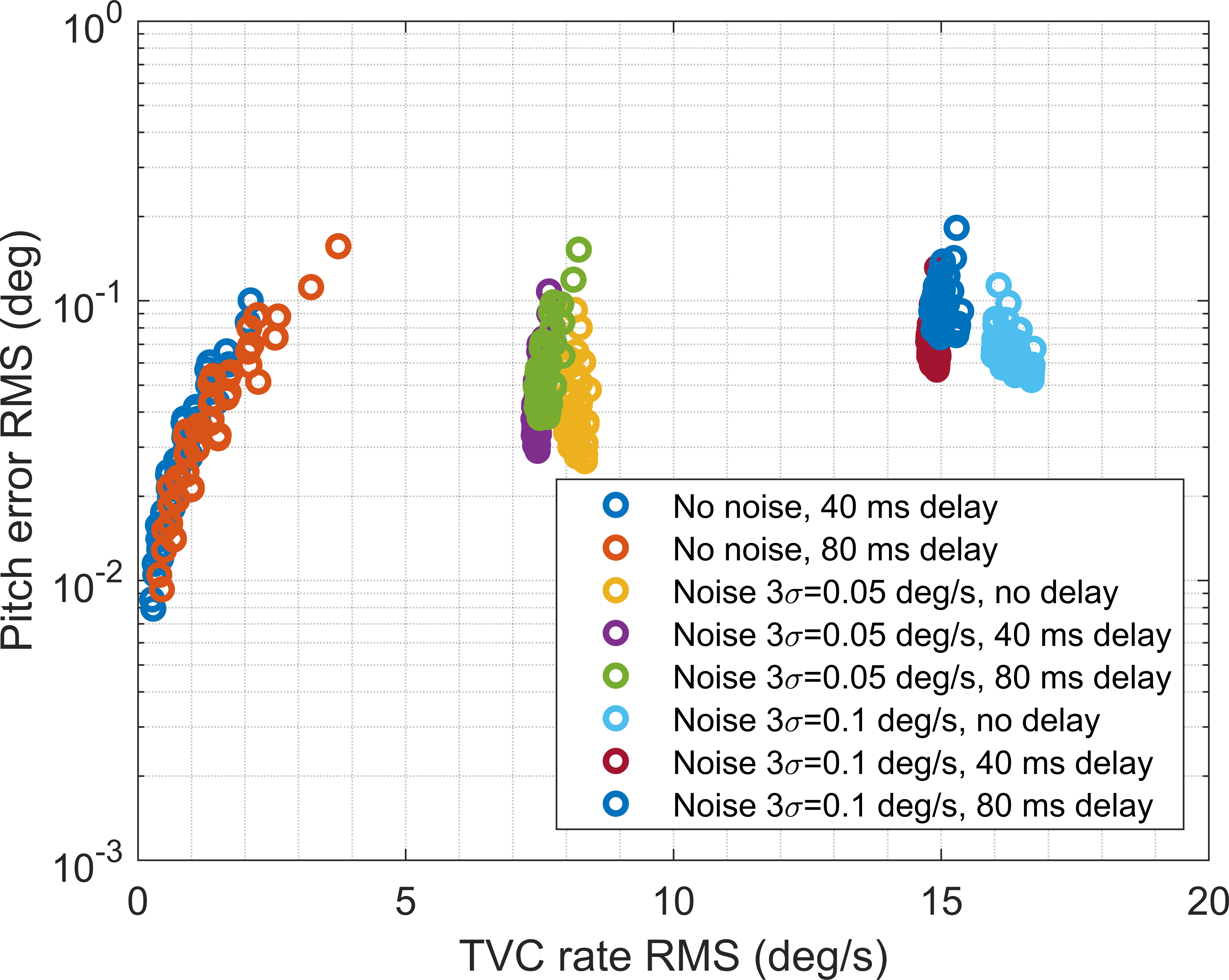}
\caption{Impact of sensor noise and actuator delays on INDI controller (w/ low-pass filter)}
\label{fig:sensitivities}
\end{figure}

INDI-based controllers, by relying on angular acceleration and control input measurements/estimates, are known to be more sensitive to sensor noise and actuator delays than classical linear controllers. In order to assess this sensitivity, Fig.~\ref{fig:sensitivities} extends Fig.~\ref{fig:results}a using the INDI controller with low-pass filter, showing wind simulation results with different combinations of:
\begin{itemize}
    \item Gaussian noise on the angular rate signal, with $3\sigma=\{0, 0.05, 0.1\}$~deg/s, which affects the estimates of both $q$ and $\dot{q}$ through Eq.~\eqref{eq:derivfilt};
    \item Time delay of \{0, 40, 80\}~ms on the signal commanded to the TVC actuator, corresponding to a delay of \{0, 1, 2\} control samples.
\end{itemize}

From Fig.~\ref{fig:sensitivities}, it can be observed that, for the ranges considered, delays on the TVC signal have very little impact on the controller's performance. Noise on the angular rate signal, on the other hand, leads to a more noticeable degradation, with the resulting TVC rates increasing approximately linearly with the noise variance. This type of understanding is therefore critical when designing and sizing INDI-based GNC software and hardware. The impact of angular rate noise would likely be minimised by using a higher-order derivative filter $H_{\dot{q}}(\mathrm{s})$ or by including an angular acceleration sensor in the GNC system.

\section{Frequency-domain robust stability analysis} \label{sec:stab}

Because of the nonlinear nature of INDI, attaining an analytical proof of stability of INDI-based controllers~\cite{wang2019indi} is much less trivial than for classical linear controllers. In order to mitigate this shortcoming, this section introduces a simple yet insightful frequency-domain approach to quantify stability degradation related to an imperfect feedback linearisation and to deviations from the control tuning conditions. This section is therefore focused on the controller developed in Sec.~\ref{sec:K4}, not on a full comparison of controllers. 

The proposed approach is based on linearised models of the nonlinear launcher simulator with the INDI control law in the loop at different flight conditions and on the fact that, for a perfect feedback linearisation, the channel $\nu\mathrm{(s)}\rightarrow\theta\mathrm{(s)}$ behaves as a double integrator (recall Eq.~\eqref{eq:linearised}). The INDI controller design was carried out under this assumption.

Linearised models of $\nu\mathrm{(s)}\rightarrow\theta\mathrm{(s)}$ can be obtained thanks to MATLAB\textsuperscript{\textregistered} routine:
\begin{center}
    \texttt{linearize(mdl, findop(mdl, $t$), ...)}
\end{center}
where \texttt{mdl} is the Simulink\textsuperscript{\textregistered} file instantiated with a certain configuration and $t$ is the flight time instant. The analysis in this section considers the $2^8=256$ corner-cases (within the uncertainty level of Table~\ref{tab:uncert}) and 33 instants (spaced every 2.5~seconds along the trajectory).

\begin{figure}[!ht]
\centering
\subfloat[Bode plot $\nu\mathrm{(s)}\rightarrow\theta\mathrm{(s)}$]
{\includegraphics[width=0.49\textwidth]{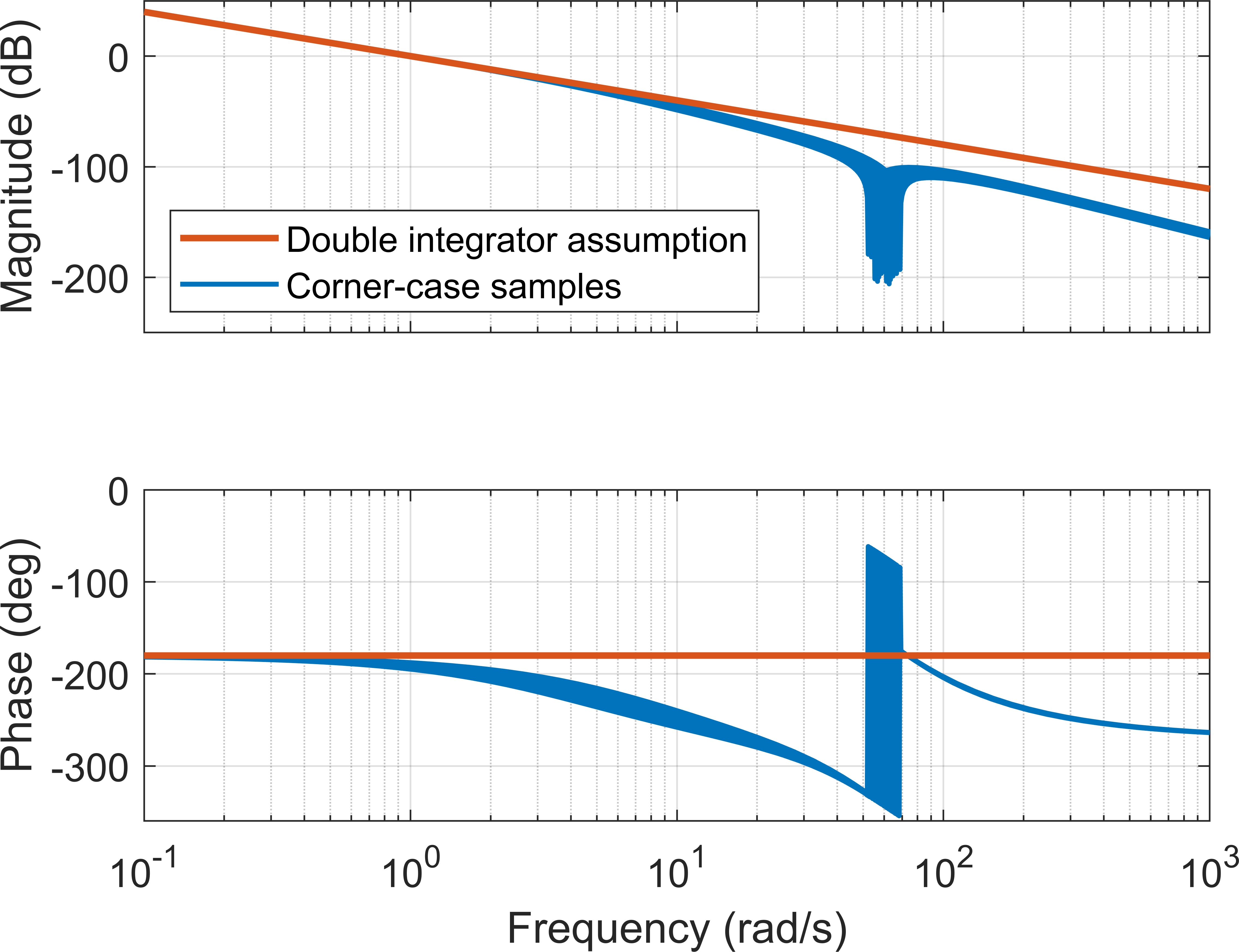}} \hfill
\subfloat[Nichols chart $\theta_{\mathrm{err}}\mathrm{(s)}\rightarrow\theta\mathrm{(s)}$]
{\includegraphics[width=0.49\textwidth]{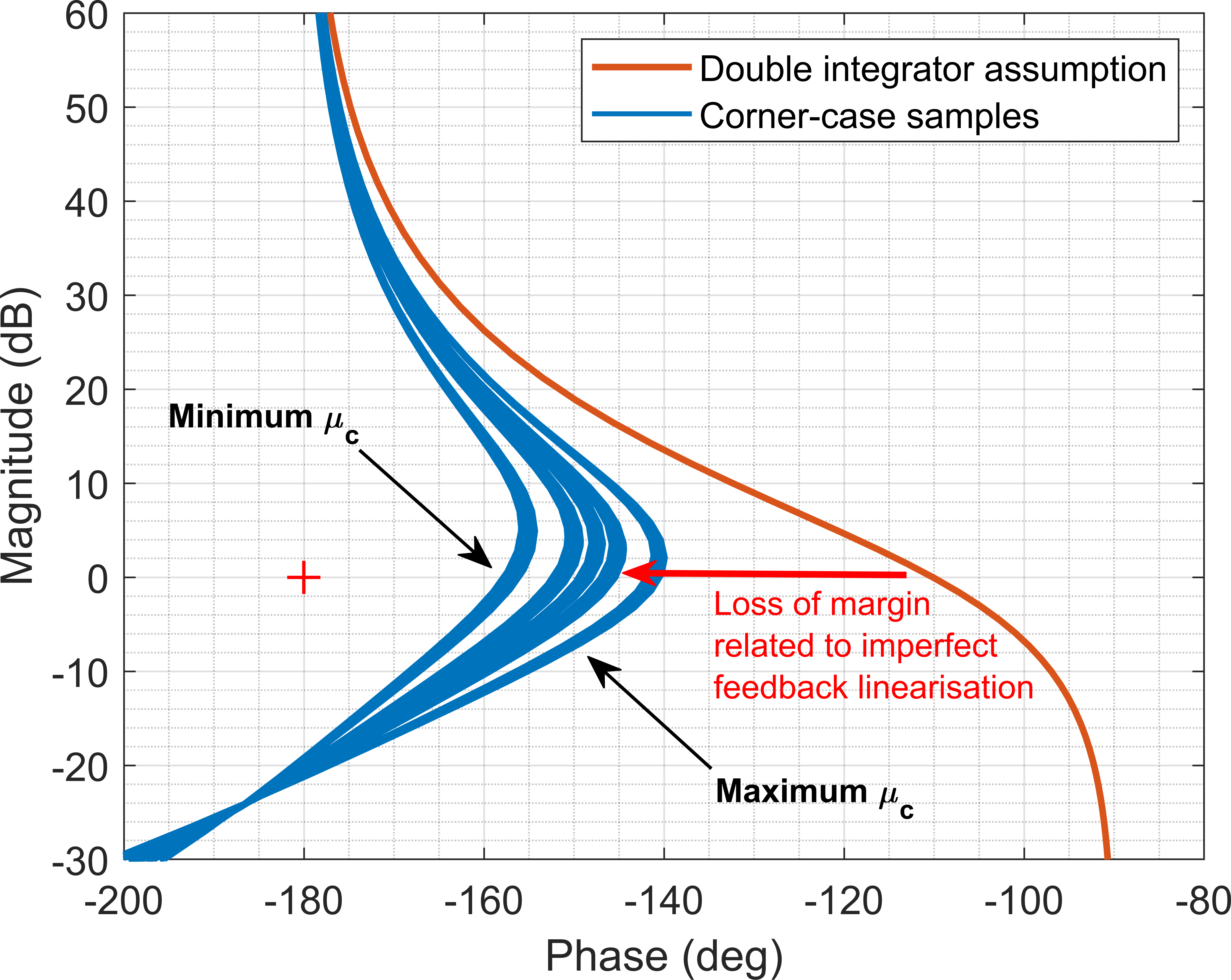}}
\caption{Frequency responses of linearised INDI-controlled plants for stability analysis}
\label{fig:stability}
\end{figure}

Figure~\ref{fig:stability}a shows the frequency response of the aforementioned linearised models (in blue), together with the "perfect" double integrator assumption (in red). This figure shows two important features:
\begin{itemize}
    \item A mismatch between the linearised models and the double integrator assumption, which grows with the frequency and arises from the fact that drift motion, TWD effects, actuator dynamics and $H_{\dot{q}}(\mathrm{s})$ filter were neglected in the feedback linearisation;
    \item A dispersion of the linearised models, which is caused by deviations from the control tuning conditions due to the uncertain and time-varying nature of the model's parameters.
\end{itemize}
These models can be employed to assess the system's stability margins when the loop is closed using Eq.~\eqref{eq:indigains}. To do so, it is convenient to plot the responses in a Nichols chart, which is depicted in Fig.~\ref{fig:stability}b. For a detailed explanation of the application of Nichols charts to launcher stability assessment, the reader is referred to~\cite{Simplicio2016}.

The impact of the imperfect feedback linearisation on the system's stability becomes evident from Fig.~\ref{fig:stability}b: the phase margin is reduced approximately by half and the system can be gain-destabilised, which is not the case under the double integrator assumption. Nonetheless, all phase and gain margins remain substantial. When this is not the case, the linearised models of $\nu\mathrm{(s)}\rightarrow\theta\mathrm{(s)}$ can be employed instead of Eq.~\eqref{eq:linearised} to re-tune the INDI outer control law. The stability margins are naturally driven by the value of $\mu_{\mathrm{c}}$, which is the main dependency of the INDI controller (recall Table~\ref{tab:depend}). Accordingly, the margins become smaller for smaller values of $\mu_{\mathrm{c}}$ as the system's control effectiveness decreases, and vice-versa.

\begin{figure}[!ht]
\centering
\subfloat[Phase margin vs. time]
{\includegraphics[width=0.49\textwidth]{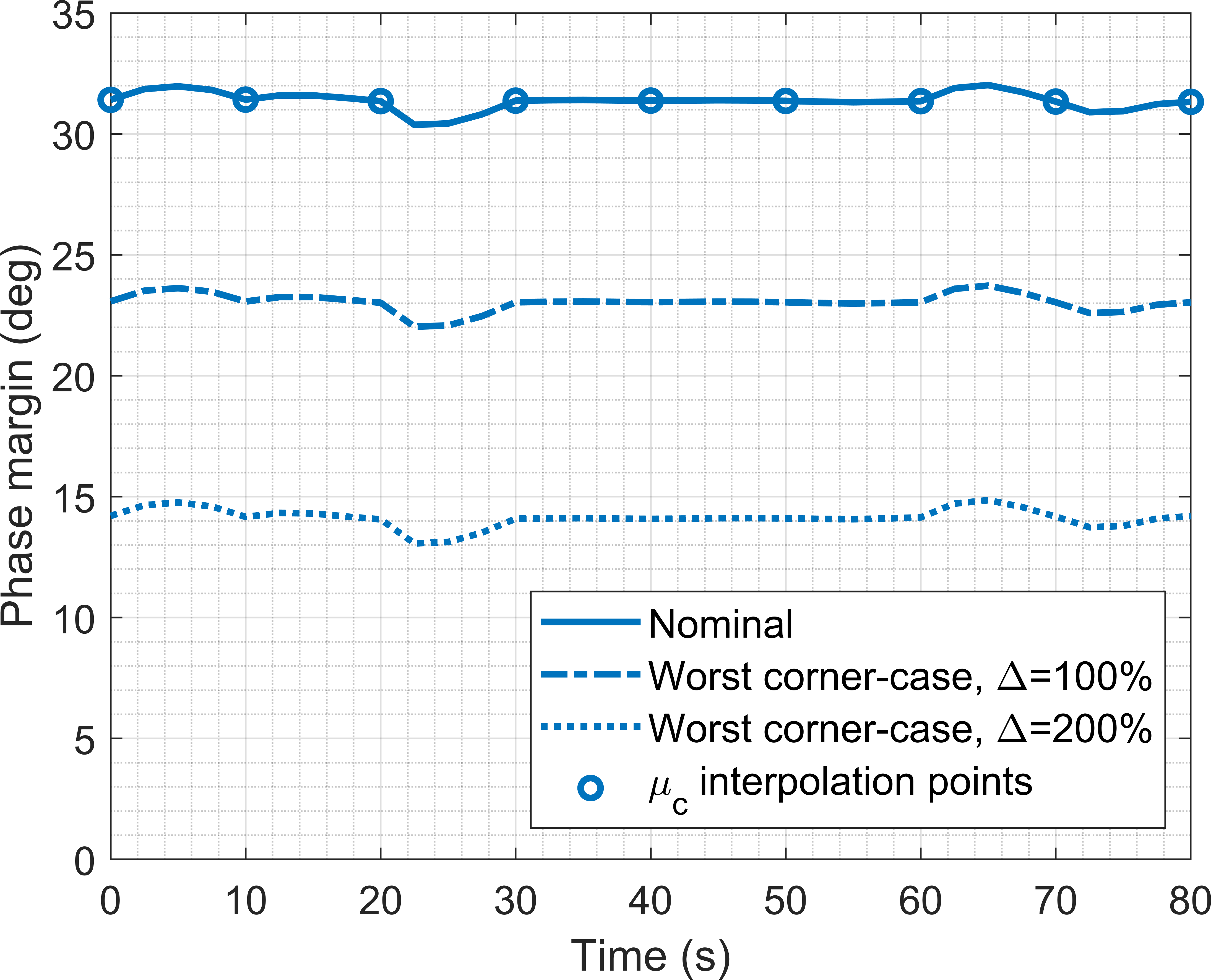}} \hfill
\subfloat[Gain margin vs. time]
{\includegraphics[width=0.49\textwidth]{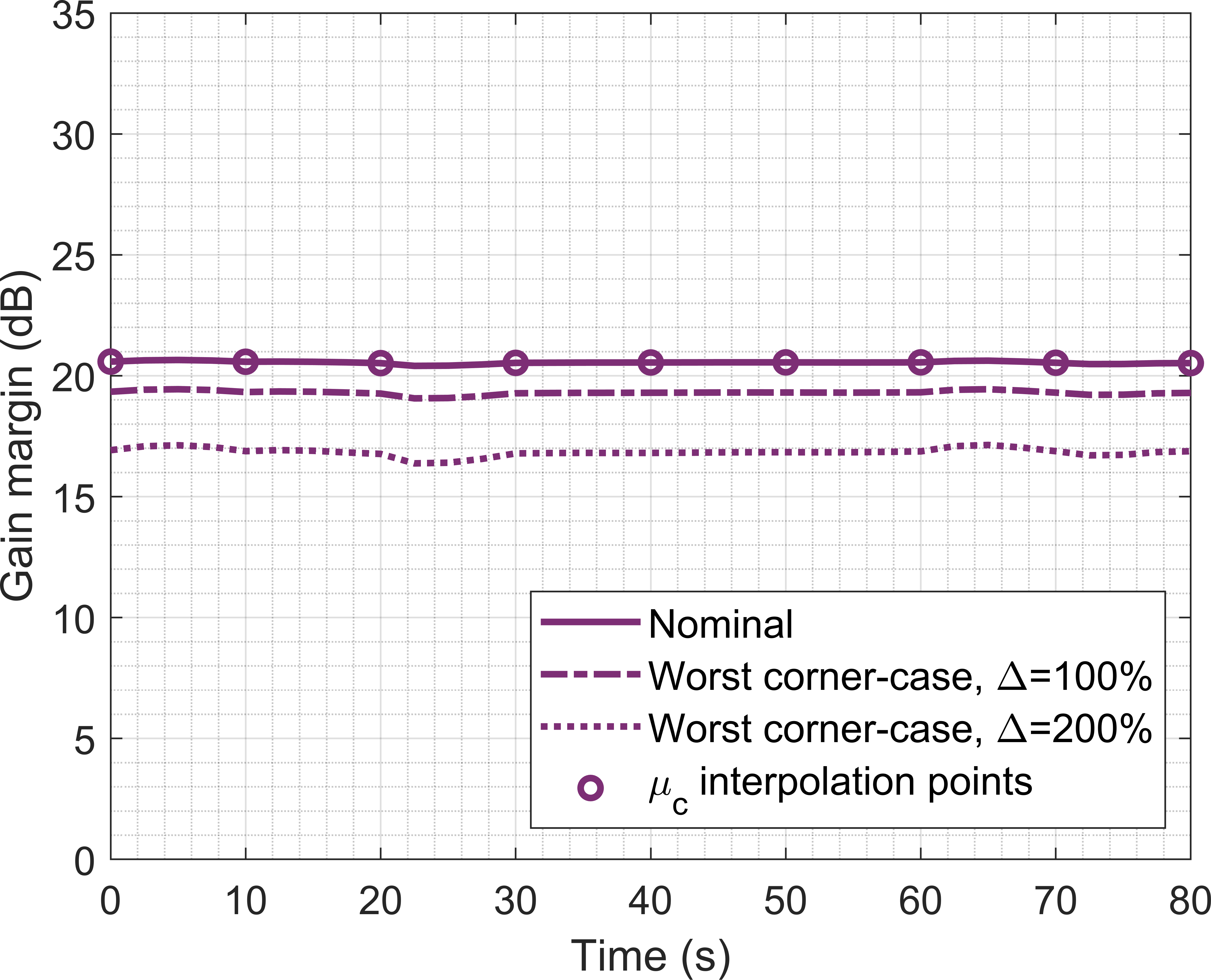}}
\caption{Nominal and worst-case margins of INDI controller (w/ low-pass filter)}
\label{fig:margins}
\end{figure}

In order to assess the degradation caused by uncertainties and time variations, the phase and gain margins are plotted as a function of time in Fig.~\ref{fig:margins}a and~b, respectively. These figures show the nominal margins (in continuous line), the worst (minimum) corner-case margins with the uncertainty level of Table~\ref{tab:uncert} ($\Delta=100\%$, in dash-dotted line) and the worst corner-case margins with twice the uncertainty level ($\Delta=200\%$, in dotted line). The $N=9$ control tuning points, i.e. the interpolation nodes of $\mu_{\mathrm{c}}$, are indicated in the figures using circular marks. The main results are then summarised in Table~\ref{tab:margins}.

\begin{table}[!ht]
\centering
\caption{Stability margin budget ($\Delta=100\%$)}
\label{tab:margins}
\vspace{-1pt}
\begin{tabular}{|l|c|c|}
\hline
\rowcolor{Gray} Case & Phase margin & Gain margin \\
\hline
Double integrator assumption & 69.84 deg & $\infty$ \\
Nominal case & 31.37 deg & 20.55 dB \\
Nominal case w/ deviation from $\mu_{\mathrm{c}}$ interpolation points & 30.38 deg & 20.41 dB \\
Worst corner-case & 23.04 deg & 19.30 dB \\
Worst corner-case w/ deviation from $\mu_{\mathrm{c}}$ interpolation points & 22.03 deg & 19.07 dB \\
\hline
\end{tabular}
\end{table}

From Fig.~\ref{fig:margins}, it can be seen that, at the control tuning points, nominal phase and gain margins are constant throughout the flight. This is expected because the closed-loop of Eq.~\eqref{eq:indigains} is time-invariant. Between tuning points there is naturally a variation in margins due to mismatches between actual and interpolated values of $\mu_{\mathrm{c}}$. Nonetheless, this variation is extremely limited and leads to a degradation of only 1~deg and 0.14~dB.

Stability degradation due to uncertainties is about one order of magnitude higher, leading to margin losses of 8.3~deg and 1.3~dB. In practice, the resulting stability margins must provide enough room to accommodate the impact of dynamical effects that were not considered in this study, such as flexible modes and non-collocated sensing. Nonetheless, the worst-case margins are plentiful, which suggests the feasibility of INDI-based launcher attitude control. In fact, the worst-case values remain acceptable even when the assumed level of uncertainty is doubled ($\Delta=200\%$, shown only in Fig.~\ref{fig:margins}, not in Table~\ref{tab:margins} for the sake of conciseness).

\section{Conclusions} \label{sec:concl}
In conclusion, this paper presented a feasibility study of  Incremental Nonlinear Dynamic Inversion (INDI) applied to a launcher ascent flight control scenario and highlighted its potential benefits over the traditional \textit{ad hoc} linear control approach widely studied and implemented in practice. The paper introduced the INDI technique which mainly cancels the nonlinearities of a (nonlinear) system by means of state/output feedback and transforms it into a \textit{linear form}, making it suitable to be controlled by a single linear control law without the need for gain-scheduling or other nonlinear approach (sliding mode, etc.). The paper also discussed the challenges associated with INDI-based control, such as sensitivity to sensor noise and actuator delay, and the difficulty of obtaining an analytical proof of stability. However, the potential benefits of INDI-based control outweighed these challenges as it was shown in a comprehensive nonlinear simulation campaign which considered wind disturbances and parameter uncertainties.

Finally, the paper proposed a simple, yet insightful linearisation-based approach to evaluate stability degradation and deviations from the (nominal) control tuning conditions. The results obtained in this study suggest that the INDI-based control approach could bring relevant improvements to launcher GNC, which may facilitate the transition to data-driven methods in the future. Outlook of this work will be furthering the analysis in terms of limits of performance (worst-case analysis) as well as addressing the impact of flexible modes and non-collocated sensing.

\section*{Acknowledgements}
The authors would like to thank Mr. Massimo Casasco, head of ESA's GNC section, for making this feasibility study possible.

\bibliographystyle{IEEEtran}
\bibliography{main}

\end{document}